\documentclass[12pt, draftclsnofoot,onecolumn,comsoc,letterpaper]{IEEEtran}
\usepackage[cmex10]{amsmath}
\usepackage{cite}
\usepackage{amssymb, mathrsfs, amsfonts, amsthm}
\usepackage{mathtools}
\usepackage[usenames,dvipsnames]{xcolor}
\usepackage{graphicx}
\usepackage{paralist}
\usepackage[colorlinks=true]{hyperref}
\usepackage{caption}
\usepackage{subcaption}
\usepackage{pgfplots}
\usepackage{tikz}
\usepackage{float}
\usetikzlibrary{plotmarks}
\usetikzlibrary{arrows.meta}
\usepgfplotslibrary{patchplots}
\usepackage{grffile}

\theoremstyle{theorem}
\newtheorem{theorem}{Theorem}
\newtheorem{lemma}{Lemma}

\newtheorem{corollary}{Corollary}
\newtheorem{definition}{Definition}
\newtheorem{example}{Example}
\newtheorem*{example*}{Example}
\newtheorem{remark}{Remark}
\newtheorem*{remark*}{Remark}

\def\IID{\mathop{\mathrm{IID}}}  
\def\crdf{\mathop{\mathrm{CRDF}}}     
\def\cldf{\mathop{\mathrm{CLF}}}     
\def\rvs{\mathop{\mathrm{RVs}}}     
\def\rv{\mathop{\mathrm{RV}}}       
\def\pmf{\mathop{\mathrm{PMF}}}     

\newcommand{\E}[1]{\mathbb{E}\left[#1\right]}
\newcommand{\Eb}[1]{\mathbb{E}\big[#1\big]}
\newcommand{\floor}[1]{\left\lfloor #1 \right\rfloor}
\newcommand{\ceil}[1]{\left\lceil #1 \right\rceil}
\renewcommand{\H}[1]{{H}\!\left( #1 \right)}
\newcommand{\Hb}[1]{{H}\big( #1 \big)}
\newcommand{\I}[2]{{I}\!\left( #1 ; #2 \right)}
\newcommand{\Ib}[2]{{I}\big( #1 ; #2 \big)}
\newcommand{\ud}{\,\mathrm{d}}

\DeclareMathOperator*{\argmax}{arg\:max}
\DeclareMathOperator*{\argmin}{arg\:min}

\allowdisplaybreaks
\allowbreak

\begin{document}
\title{Secure Block Source Coding with Sequential Encoding}
\author{Hamid Ghourchian, \IEEEmembership{Student Member, IEEE},
	Photios A. Stavrou, \IEEEmembership{Member, IEEE}
	Tobias J. Oechtering, \IEEEmembership{Senoir Member, IEEE}, and
	Mikael Skoglund, \IEEEmembership{Fellow, IEEE}
	\thanks{This work was supported in part by the Swedish Foundation for Strategic Research, KTH Digital Futures, and SRA ICT-TNG.}
	\thanks{Part of this work (without security constraints) was presented in IEEE Information Theory Workshop, 2019 \cite{hamid19}.}
	\thanks{The authors are with the Department of Intelligent Systems, Division of Information Science and Engineering, at KTH Royal Institute of Technology, 10044 Stockholm, Sweden (e-mail: \mbox{\{hamidgh; fstavrou; oech; skoglund\}@kth.se}).}}

\date{}
\maketitle

\begin{abstract}
We introduce fundamental bounds on achievable \emph{cumulative rate distribution functions} ($\crdf$) to characterize a sequential encoding process that ensures lossless or lossy reconstruction subject to an average distortion criterion using a non-causal decoder.
The $\crdf$ describes the rate resources spent sequentially to compress the sequence.
We also include a security constraint that affects the set of achievable $\crdf$.
The information leakage is defined sequentially based on the mutual information between the source and its compressed representation, as it evolves. 
To characterize the security constraints, we introduce the concept of \emph{cumulative leakage distribution functions} ($\cldf$), which determines the allowed information leakage as distributed over encoded sub-blocks.
Utilizing tools from majorization theory, we derive necessary and sufficient conditions on the achievable $\crdf$ for a given independent and identically distributed ($\IID$) source and $\cldf$.
One primary result of this paper is that the concave-hull of the $\crdf$ characterizes the optimal achievable rate distribution.
\end{abstract}


\section{Introduction}
\IEEEPARstart{I}{n} this paper, we consider a source coding problem with sequential encoding where a memoryless and independent and identically distributed ($\IID$) source is communicated in chunks to a decoder, which uses lossless or lossy reconstruction, subject to a single-letter distortion constraint.
The coding is done sequentially in sub-blocks, i.e., the whole source sequence is split into sub-blocks, which are sequentially observed by the encoder, while the decoder decodes all the sub-blocks at once.
Hence, the encoding is causal in the sense that the encoder does not access future sub-blocks, whereas a joint decoding of all sub-blocks is considered for the decoder.
We also consider security constraints on the sequential encoding to prevent the encoder from revealing too much information about the part of the $\IID$ source sequence that corresponds to the messages sent so far.

Sequential encoding the way we introduce it in this paper can be motivated by complexity, delay, or channel availability constraints.
Consider, for instance, a system with low-end hardware for the encoder side so the encoder can only afford to process a limited amount of source information at a time because of storage or computational constraints.
There may also be a delay constraint at the encoder side, in the sense that the encoder has to deliver a message within a specific time range.
This latter scenario can be the case, for example, when the encoder works over a channel that is available only sporadically and at certain moments in time.
Given available scheduling of the communication channel, corresponding to a particular \emph{cumulative rate distribution function} ($\crdf$; see Definition \ref{def:CRDF}), our new theory can tell whether this particular scheduling of the channel is achievable for source coding at a specified fidelity.
Practical scenarios for such scheduling problems include the case of a vehicle communicating time-series of measurements to the background radio access network following a resource management strategy taking varying coverage into account.

The information leakage constraint we introduce herein, corresponding to a given \emph{cumulative rate leakage function} ($\cldf$; see Definition~\ref{def:CLDF}), is motivated by the scenario where side information is available regarding the presence of an eavesdropper listening to the messages transmitted from the encoder so far.
Alternatively, it can be motivated by a scenario where an eavesdropper is present throughout the transmission, and we wish to reveal information only according to the specific allowed patterns motivated by the application at hand.
For example, in a sensor network, the source sequence can correspond to a long time-series of measurements to be communicated in sub-blocks at low cost to a central processing node.
Such communication can happen over a relatively long time (hours or even days); so it may be beneficial to reveal the measurements, as little as possible, during the earlier times while allowing higher data leakage when the whole process is close to being finalized.
Such a scenario can be motivated by not allowing a passive adversary to draw early conclusions that can pre-date the final decision made at the central processing node.

One major result of this paper is that the achievable $\crdf$ subject to a certain given $\cldf$ reveals a fundamental rate allocation bound that depends only on the concave-hull of the rate profile.
This result is useful in practice since it simplifies the design space for the optimization of resource allocation policies.

\emph{\bf Literature Review}:
Classical rate-distortion theory was introduced in \cite{Shannon59} and characterizes the fundamental trade-off between the achievable distortion and the rate of a non-causal encoder and decoder pair. 
An excellent overview of the classical results can be found, for instance, in \cite{berger:1971}. 
A variant of classical rate-distortion function called the OPTA by causal codes was introduced in \cite{Neuhoff82}. 
In that framework, reconstruction of the present source sample is restricted to be a function of the present and past source samples, while the code stream itself may be non-causal and have a variable rate. 
A generalization of \cite{Neuhoff82} when the framework is allowed to have side information can be found in \cite{Weissman05}. 
A subclass of causal source coding is zero-delay coding, where the encoder and decoder operate instantaneously, see, for example, \cite{Linder01}. 
Recently, causal and zero-delay source coding received particular attention by both information theorists and the control community.
The reason is that such compression schemes appear to be appropriate to derive fundamental performance limitations in closed-loop control systems, see, for example, \cite{Stavrou18,Tanaka18} and references therein.
Causal and zero-delay source coding were also used in the context of source-channel coding applications, or source coding with finite memory, see, for example, \cite{Akyol14, Matloub06, Merhav03}. 
Sequential source coding \cite{Viswanathan00, Ma11} is also another kind of source coding, and it is similar to our framework when the number of the encoders tends to infinity.
However, in contrast to that coding paradigm that aims to characterize the rate-region of finite encoders, here, besides the security constraint, we also characterize the rate profile of all the encoders when their number tends to infinity.

Shannon originally introduced the notion of security from an information-theoretical perspective in \cite{Shannon49}.
A few decades later, Wyner introduced the celebrated wiretap channel \cite{Wyner75} and showed that it is possible to send information at a positive rate with perfect secrecy when eavesdropper's channel is a degraded version of the channel from the encoder to the decoder.
When it comes to secure communication via information-theoretic tools, often, two approaches are encountered in the literature.
The first one presupposes that both encoder and decoder agree on a secret key before the transmission of the source.
The second approach assumes that the decoder and the eavesdropper (sometimes the encoder as well) have different versions of side information, and thereby secrecy is achieved through this difference.
For instance, Shannon, in \cite{Shannon49}, using the first approach, showed that the transmission of a discrete memoryless source is entirely secure if the rate of the key is at least as large as the entropy of the source. 
Yamamoto in \cite{Yamamoto97} studied various secure source coding scenarios that include, among other results, an extension of Shannon's cipher system to combine secrecy with rate-distortion theory.
Prabhakaran and Ramchandran in \cite{Prabhakaran07} considered lossless source coding with side information at both the decoder and the eavesdropper when there is no rate constraint between the encoder and the decoder.
In \cite{Gunduz08}, the authors considered a setup with side information at the encoder and coded side information at the decoder.
Villard and Piantanida in \cite{Villard13} extended these works.
The authors therein studied the problem of secure lossy source coding when one or both the receiver and the eavesdropper have side information.
In \cite{Schieler14}, the authors considered secrecy in communication systems by the distortion that an adversary incurs.
In their setups, both the transmitter and receiver share a secret key, which is used to encrypt communication and ensure distortion at the adversary.
Kaspi and Merhav in \cite{Kaspi15} considered two source coding models combining causal or zero-delay source coding with secrecy constraints.

Majorization theory has been extensively used in communications and information theory. 
For instance, in \cite{Shamai03}, it was used to derive a broadcast approach for a single-user slow fading MIMO channel, and in \cite{Jorswieck07, Palomar07}, it was used in the context of optimal rate allocation and transceiver design of vector-valued wireless communication systems.

This paper is structured as follows.
In Section \ref{sec:ProblemStatement}, we formally introduce $\crdf$, $\cldf$, and our sequential source coding problems.
We give our main results in Section \ref{sec:main_results}.
In Section \ref{sec:proofs}, we provide the derivations of the proofs. 
In Section \ref{sec:usefulLemmas}, we give some lemmas, which are utilized in the derivations of our main results.
Finally, we draw conclusions in Section \ref{sec:conclusion}.

\emph{Notations}:
Sets, random variables ($\rvs$) and their realizations are denoted by calligraphic, capital and lower case letters, respectively.
The set of integer, rational and real numbers are denoted by $\mathbb{N}$, $\mathbb{Q}$ and $\mathbb{R}$, respectively.
The probability mass function ($\pmf$) of a random variable $X$ with realizations $X=x$ defined on some alphabet $\mathcal{X}$ of finite cardinality $\lvert\mathcal{X}\rvert$ is denoted by $p_X(x)$ or just $p(x)$.
Similarly, for two $\rvs$ $X$ and $Y$, the conditional $\pmf$ of $Y$ given $X=x$ is denoted by $p_{Y|X}(y|x)$ or just $p(y|x)$.
The sequence $(x_{m}, x_{m+1}, \ldots, x_n)$, for $m,n \in\mathbb{N}$, is denoted by $x_{m}^{n}$.
If $m=1$, we may use the notation $x^n$ instead of $x_1^n$.
Also, $x^0$ means $\emptyset$.
The notation $\mathbb{E}[X]$ means the expected value of $\rv$ $X$.
All logarithms are in base $2$ unless otherwise stated.
The term ``w.r.t.'' is an abbreviation for ``with respect to''.

\section{Problem Statement and New Definitions} \label{sec:ProblemStatement} 
\begin{figure}[h]
	\centering
	\includegraphics[trim = 0 250 0 0, clip, width=\columnwidth, scale=1]{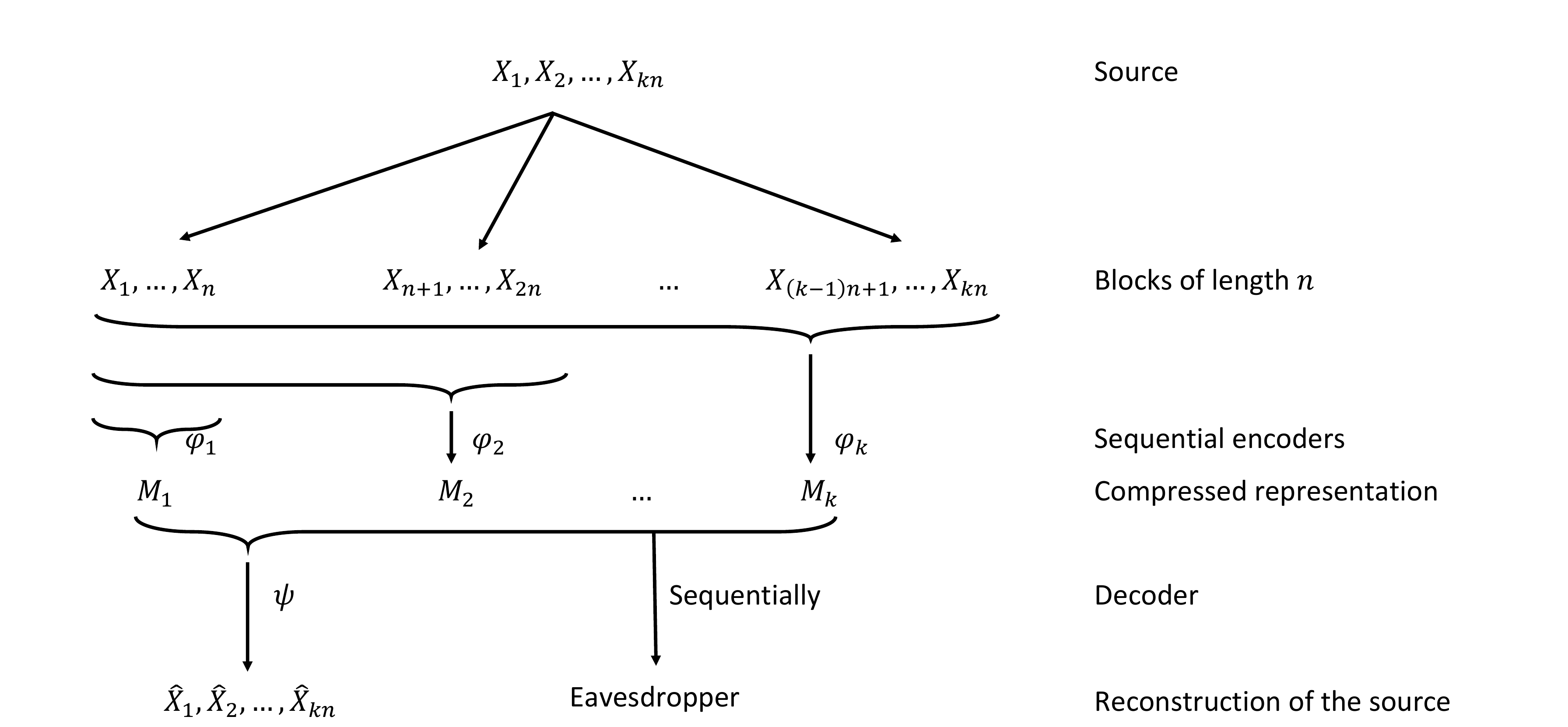}
	\caption{The encoders $\varphi_1,\ldots,\varphi_k$ sequentially encode the current and past sub-blocks of length $n$.
	The rate of the encoders specifies the rate profile. 
	The decoder $\psi$ jointly decodes all messages $M_1,\dots M_k$ at once.
	The eavesdropper sees the messages sequentially as they are transmitted.
	\label{fig:system}}
\end{figure}

In this section, we define our problem formulation and some new definitions.
As illustrated in Fig.~\ref{fig:system}, the source block is a sequence of $\IID$ $\rvs$ of length $nk$, which is divided into $k$ sub-blocks of the length of $n$ random variables.
For each sub-block, there is an encoder that has access to the source symbols of all the previous and the current sub-blocks but not the future ones, corresponding to (block) causal encoding.
The output of the $i$-th encoder is denoted by $M_i$.
However, we assume that the decoder is not constrained to be causal, i.e., it can wait until it has received all messages $M_1,\ldots,M_k$ corresponding to the whole source sequence.
The allocated rates to encode the sub-blocks follow a certain $\crdf$ (see Definition~\ref{def:CRDF}).
Besides, we also enforce a $\cldf$ (see Definition~\ref{def:CLDF}), motivated by the potential presence of an eavesdropper who can overhear the transmitted messages of the encoded sub-blocks.
The information leakage is measured sequentially as the encoding of sub-blocks progresses.
The leakage constraint is formulated in terms of the revealed mutual information over time (see Definitions~\ref{def:Achievable Lossless Codes}, \ref{def:Achievable R(D)}).

\begin{definition}[Regular Cumulative Function] \label{def:Regular Cumulative Function}
	A function $F \colon [0,1] \to [0,\infty)$ is a \emph{regular cumulative function} if it satisfies the following properties:
	\begin{enumerate}
		\item
		Cumulation: $F$ is non-decreasing,
		\item
		Zero initial value: $F(0) = 0$,
		\item
		Regularization: $F$ is continuous from the right, i.e.,
		$\lim_{\beta\searrow\alpha}F(\beta) = F(\alpha)$ for all $\alpha\in [0,1)$.
	\end{enumerate}
\end{definition}
Regular cumulative functions represent both the cumulative rate at which the encoding is allowed and the allowable leakage.

\begin{definition}[Cumulative Rate Distribution Function] \label{def:CRDF}
	A function $G \colon [0,1] \to [0,\infty)$ is a $\crdf$ if $G$ is a \emph{regular cumulative function} characterizing the cumulative rate at which encoding is allowed.
	The domain of the function $G$ represents the normalized time of blocks for the transmission of the whole sequence of blocks; as a result, if there are $k$ blocks, $G(\alpha)$ represents the accumulated rate for compression until block $\floor{\alpha k}$ ends.
\end{definition}

Because the function characterizes the cumulative rate, it is non-decreasing.
There is no need to consider an available rate before the start of the sequence; so, it has zero initial value.

\begin{definition}[Cumulative Leakage Function] \label{def:CLDF}
	A function $L \colon [0,1] \to [0,\infty)$ is a $\cldf$ if $L$ is a \emph{regular cumulative function} characterizing the cumulative leakage constraint over time.
	The domain of the function $L$ represents the normalized time of blocks w.r.t. the whole time of the sequence; as a result, if there are $k$ blocks of $n$ symbols, $L(\alpha)$ represents the allowable leakage of the compressed messages until block $\floor{\alpha k}$ ends about the $n \floor{\alpha k}$ first symbols of the source, i.e., the leakage of $M^{\floor{\alpha k}}$ about $X^{n\floor{\alpha k}}$.
\end{definition}

We assume that the leakage is not reduced when time passes. 
It is also consistent with the definition of mutual information as the leakage (see Definitions \ref{def:Achievable Lossless Codes}, \ref{def:Achievable R(D)}), i.e., $\I{X^{ni}}{M^i}\geq \Ib{X^{n(i-1)}}{M^{i-1}}$.
Hence, the leakage function is non-decreasing.
Similar to $\crdf$, the leakage function has zero initial value.

Note that for a fixed number of sub-blocks $k$, as $n\to\infty$, the samples of $\crdf$ and $\cldf$, at the points $i/k$ for $i=1,\ldots,k$, determine the rate and the leakage profiles, respectively.
As a result, only those points are required.
The other points of the functions become necessary, as $k$ increases. 
Thus, for a fixed $k$, it is the same as evaluating the rate and leakage profiles according to the step-wise functions $\alpha\mapsto G(\floor{\alpha k} / k)$ or $\alpha\mapsto L(\floor{\alpha k} / k)$, respectively.

\begin{definition}[Sequential encoding with $(G, k, n)$-source codes] \label{def:Code}
	Assume $(X_1, X_2, \ldots, X_{n k})$ is a block of $\rvs$ each one defined on domain $\mathcal{X}$.
	A $(G, k, n)$-source code, where $(k, n)\in\mathbb{N}$, and $G \colon [0,1] \to [0,\infty)$ is a $\crdf$ according to Definition \ref{def:CRDF}, consists of
	\begin{itemize}
		\item an ensemble of $k$ sequential encoders $\varphi_1,\ldots,\varphi_k$ such that each one of them assigns an index to the source sequence blocks received so far, i.e., for each $i\in\{1,\ldots,k\}$,
		\begin{equation*}
			\varphi_i \colon \mathcal{X}^{i n} \to \mathcal{M}_i,
			\quad x^{i n} \mapsto m_i,
			\qquad\mathcal{M}_i := \left\{ 1, \ldots, 2^{\floor{n k R_i}} \right\}, 
			\quad i=1,\ldots,k,
		\end{equation*}
		where
		\begin{equation} \label{eqn:def R}
			R_i = G\left( \frac{i}{k} \right) - G\left( \frac{i-1}{k}\right),
			\qquad i=1,\ldots,k,
		\end{equation}
		\item a decoder, $\psi$, that reconstructs $\hat{x}^{n k}$ based on the output of the encoders, i.e.,
		\begin{equation*}
			\psi\colon \otimes_{i=1}^k{\mathcal{M}_i} \to \hat{\mathcal{X}}^{n k},
			\quad (m_1,\ldots,m_k) \mapsto \hat{x}^{n k},
		\end{equation*}
		where $\otimes$ denotes the Cartesian product and $\hat{\mathcal{X}}$ denotes the reconstruction domain.
	\end{itemize}
\end{definition}

Next, we introduce the definition of the achievable $\crdf$s for both lossless and lossy compression.
\begin{definition}[Achievable $\cldf$-secure $\crdf$ for lossless compression] \label{def:Achievable Lossless Codes}
	Assume that $X_1, X_2, \ldots$ is a sequence of $\rvs$, each with support $\mathcal{X}$.
	A $\crdf$ $G$ is said to be achievable to encode the sequence $X_1, X_2, \ldots$, given $\cldf$ $L$, in a lossless manner, if for any $k\in\mathbb{N}$, there exists a sequence of $(G, k, n)$-source codes for $n\in\mathbb{N}$, with output domain $\hat{\mathcal{X}}=\mathcal{X}$, such that
	\begin{IEEEeqnarray}{l}
		\lim_{n\to\infty}{\Pr\{X^{nk} \neq \hat{X}^{nk}\}} = 0, \label{eqn:Pe->0}\\
		\frac{1}{n k}\Ib{X^{i n}}{M^i} \leq L\left(\frac{i}{k}\right),
		\qquad \forall i\in\{1, 2, \ldots, k\}, \label{eqn:I<L lossless}
	\end{IEEEeqnarray}
	where \eqref{eqn:Pe->0} denotes the probability of error that tends to zero for any $k\in\mathbb{N}$ as $n\to\infty$;
	\eqref{eqn:I<L lossless} denotes the normalized amount of leakage of the first $i$ blocks, while the messages until the end of block $i$ have been measured.
\end{definition}

Note that, there exist some $\cldf$ with no achievable $\crdf$.
For example, for $\IID$ $\rvs$, at least $1/(nk) \H{M^k} \geq \H{X}$ bits must be disclosed at the end of the whole block to satisfy \eqref{eqn:Pe->0} (due to the classical lossless source coding \cite[Theorem 3.4]{ElGamal:2011}). 
Hence, if for a given $\cldf$ $L$, we have $L(1) < \H{X}$, then there is no feasible $\crdf$ with $\cldf$ $L$.

\begin{definition}[Achievable $\cldf$-secure $\crdf$ for lossy compression] \label{def:Achievable R(D)}
	Assume that $X_1, X_2, \ldots$ is a sequence of $\rvs$, each with support $\mathcal{X}$.
	A $\crdf$ $G$ is said to be achievable to encode the sequence $X_1, X_2, \ldots$, given $\cldf$ $L$ and an average expected distortion level less than $\bar{d}$, if for any $k\in\mathbb{N}$, there exists a sequence of $(G, k, n)$-source codes, for $n\in\mathbb{N}$, with output support $\hat{\mathcal{X}}$, such that
	\begin{IEEEeqnarray}{l}
		\limsup_{n\to\infty}{\E{d(X^{n k},\hat{X}^{n k})}}
		\leq \bar{d}, \label{eqn:limsupEd<bard}\\
		\frac{1}{n k}\Ib{X^{i n}}{M^i} \leq L\left(\frac{i}{k}\right),
		\qquad \forall i\in\{1, 2, \ldots, k\}, \label{eqn:mainI<L}
	\end{IEEEeqnarray}
	where for the given distortion function $d \colon \mathcal{X}\times\hat{\mathcal{X}} \to [0,\infty]$, we have
	\begin{equation*}
		d(X^{n k},\hat{X}^{n k})
		:= \frac{1}{n k} \sum_{i=1}^{n k}{d(X_i,\hat{X}_i)}.
	\end{equation*}
	Similar to \eqref{eqn:I<L lossless}, \eqref{eqn:mainI<L} denotes the normalized amount of leakage of the first $i$ blocks, while the messages until the end of block $i$ have been measured.
\end{definition}

\section{Main Results} \label{sec:main_results}
In this section, we characterize the set of all achievable $\crdf$s for both lossless and lossy compression.
We assume that the sequence of $\rvs$ is $\IID$ defined on a finite support domain.
The essence of our sequential approach lies in that the rates available later can be used to compress previous source symbols; thus, the required rate to compress a sub-block can be split, and some parts can be sent later.
To make this point clear, consider a rate profile $\{R_i^{(1)}\}_{i=1}^k$ that majorizes (for details on majorization, see Lemma \ref{lmm:MajIneq}) another rate profile $\{R_i^{(2)}\}_{i=1}^k$, i.e.,
\begin{equation*}
	\begin{cases}
		\sum_{i=1}^j R_i^{(2)}
		\leq \sum_{i=1}^j R_i^{(1)},
		& j = 1,\ldots,k-1, \\
		\sum_{i=1}^k R_i^{(2)}
		= \sum_{i=1}^k R_i^{(1)}.
		& ~
	\end{cases}
\end{equation*}
Then the achievability of $\{R_i^{(1)}\}_{i=1}^k$ implies the achievability of $\{R_i^{(2)}\}_{i=1}^k$ (see Lemma \ref{lmm:CRDFMaj} for more details).
Further, utilizing the fact that the rate-distortion function is convex, we use the majorization inequality (see Lemma \ref{lmm:MajIneq}) to show that if a rate profile is achievable, its concave-hull is also achievable.

The main idea behind the security constraint is that the encoder may not be allowed to use all the available rates at any time because more than allowed information would be revealed to the eavesdropper.
Hence, the encoder must send some of the information later. 
Therefore, a given rate profile may not be effective in the sense that an amount of the allowable rate in some blocks must not be used to avoid revealing too much information. 
As a result, an effective $\crdf$ is defined to identify the efficient amount of rate that can be used in the coding scheme with security constraints. 

The following two theorems are the main results of this paper.
\begin{theorem}[Achievable collection of $\crdf$s via lossless compression] \label{thm:lossless}
	For an $\IID$ sequence $X_1,X_2,\ldots$, functions $G$ and $L$ are $\crdf$ and $\cldf$, respectively. 
Then, $G$ is achievable, with $\cldf$ $L$, in the sense of Definition \ref{def:Achievable Lossless Codes}, if and only if
	\begin{equation} \label{eqn:F1-F0<H}
		G(1) - G(\alpha)
		\geq \max\{(1-\alpha) \H{X}, \H{X} - L(\alpha) \},
		\qquad \forall \alpha\in[0,1],
	\end{equation}
	where $X$ is a $\rv$ with $\pmf$ $p(x)$ over the finite domain $\mathcal{X}$.
\end{theorem}
\begin{IEEEproof}
	See Section~\ref{subsec:prf:thm:lossless}.
\end{IEEEproof}

Note that, for $\alpha=1$ in \eqref{eqn:F1-F0<H},
\begin{equation*}
	0 
	= G(1) - G(1)
	\geq H(X) - L(1)
	\Longrightarrow
	L(1) \geq \H{X},
\end{equation*}
which is consistent with the discussion after Definition \ref{def:Achievable Lossless Codes}.
Further, for any fixed number of sub-blocks, $k$, as $n\to\infty$, only the values of $G(i/k)$ and $L(i/k)$ for $i=1,\ldots,k$, are important, according to Definitions \ref{def:Code} and \ref{def:Achievable Lossless Codes}.
Hence, it can be shown that \eqref{eqn:F1-F0<H} must be correct only in $\alpha = i/k$ for $i=1,\ldots,k$.
However, when $k$ becomes large, the value of the other points becomes necessary, and \eqref{eqn:F1-F0<H} is obtained for all $\alpha\in[0,1]$.
	
The idea on which the proof of Theorem \ref{thm:lossless} is based is that the amount of bits per symbol used to encode each block must be at least $\H{X}$.
Hence, there is no need to use more than $\H{X}$ bits per symbol to compress the sequence in total. 
Therefore, we remove the rates of the first blocks, because the rates of the next blocks can compensate them (follows from Lemma \ref{lmm:CRDFMaj}).
We define effective $\crdf$ $\bar{G}^{\rm eff}$ as
\begin{equation} \label{eqn:barGeff}
	\bar{G}^{\rm eff}(\alpha)
	:= \max\{0, G(\alpha) - (G(1) - \H{X})\}.
\end{equation}
The security constraint is also satisfied because $\bar{G}^{\rm eff}(\alpha) \leq L(\alpha)$ which follows from \eqref{eqn:F1-F0<H} (see the proof for the details).
Hence, it should be possible to shift the rates of $\bar{G}^{\rm eff}$ such that $\H{X}$ bits per symbol are allocated to each block.
Thus, the theorem follows.

In order to state the next theorem, we first need the following definitions.
\begin{definition}[Concave-hull or envelope of a function] \cite[p. 119]{Boyd04} \label{def:ConvHull}
	Let $f\colon \mathcal{A} \to \mathbb{R}$ be a function with a convex domain $\mathcal{A}$.
	Then, $\hat{f}\colon \mathcal{A} \to \mathbb{R}$ is the \emph{concave hull or envelope} of $f$ if $\hat{f}$ is a concave function such that $f(x) \leq \hat{f}(x),~\forall x\in \mathcal{A}$, and for any concave function $g \colon \mathcal{A} \to \mathbb{R}$ such that $f(x) \leq g(x),~\forall x\in \mathcal{A}$, we have $\hat{f}(x) \leq g(x),~\forall x\in \mathcal{A}$.
\end{definition}

\begin{definition}[Rate-distortion and distortion-rate functions] \cite[p. 307]{Cover06} \label{def:classical R(D)}
	For a given distortion function $d \colon \mathcal{X}\times\hat{\mathcal{X}} \to [0,\infty)$ and a probability distribution $X\sim p(x)$, the rate distortion function, $R(D)$, is defined based on \cite[p. 307]{Cover06}.
	The inverse of $R(D)$ is the distortion-rate function, denoted by $D(R)$, such that $D(R) = \min\{D \colon R(D) = R\}$.
\end{definition}

The $R(D)$ and $D(R)$ satisfy well-known functional and topological properties (see, for example \cite{berger:1971,Cover06}).
In the next remark, we state some of the most important of them as these properties used in the derivation of our main result.
\begin{remark} \label{cor:R(D)convex}
	$R(D)$ and $D(R)$ are non-increasing and convex functions of $D\in[0,\infty)$ and $R\in[0,\infty)$, respectively. 
	Besides, $R(D)$ and $D(R)$ are continuous w.r.t. $D\in(0,\infty)$ and $R\in(0,\infty)$, respectively.
	Further, if $R(0)<\infty$ or $D(0)<\infty$, then it is continuous w.r.t. $D\in[0,\infty)$ or $R\in[0,\infty)$, respectively.
\end{remark}

\begin{theorem}[Achievable region via lossy compression] \label{thm:R(D)}
	Assume $X_1, X_2, \ldots$ is an $\IID$ sequence with $\pmf$ $p(x)$ and the finite support $\mathcal{X}$.
	Further, assume that a distortion function $d \colon \mathcal{X}\times\hat{\mathcal{X}} \to [0,\infty]$ is given and we have that the distortion-rate function, $D\colon [0,\infty) \to [0,\infty)$, (see Definition \ref{def:classical R(D)}) is bounded, i.e., $\sup_{R\geq 0}D(R) < \infty$.
	Define $\crdf$ $G^{\rm eff}\colon [0,1] \to [0,\infty)$ as
	\begin{equation} \label{eqn:GL}
		G^{\rm eff}(\alpha)
		:= \max\left\{0, G(\alpha) - \sup_{\beta\in[0,1]}(G(\beta) - L(\beta))\right\}.
	\end{equation}
	Then, the $\crdf$ $G$ is achievable, given $\cldf$ $L$, with distortion level $\bar{d}$, in the sense of Definition \ref{def:Achievable R(D)}, if and only if
	\begin{equation} \label{eqn:intDG<d}
		\int_0^1 {D\left(\frac{\ud \hat{G}^{\rm eff}}{\ud \alpha}(\alpha) \right) \ud \alpha}
		\leq \bar{d},
	\end{equation}
	where $\hat{G}^{\rm eff} \colon [0,1] \to [0,\infty)$ is the envelope of the function $G^{\rm eff}$ in the sense of Definition \ref{def:ConvHull} and $D(\cdot)$ is the distortion-rate function.
\end{theorem}

\begin{IEEEproof}
	See Section \ref{subsec:prf:R(D)}.
\end{IEEEproof}

\begin{figure}
\centering
	\input{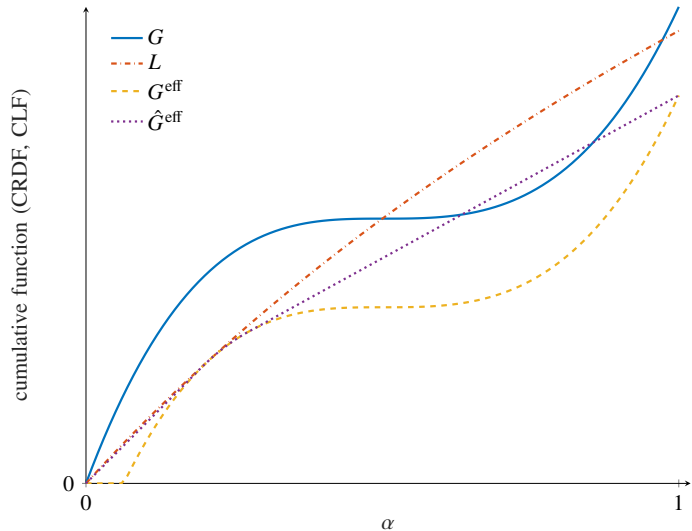}
	\caption{An illustration of Theorem \ref{thm:R(D)}.}
	\label{fig:thm:R(D)}
\end{figure}

In Fig. \ref{fig:thm:R(D)} we illustrate an example of the variables used in Theorem \ref{thm:R(D)}.

	The idea behind Theorem \ref{thm:R(D)} is that the minimum possible amount of distortion of each block, with $\tilde{R}$ available bits, is $D(k\tilde{R})$.
	The amount of the leaked information is $\tilde{R}$ bits, which must be less than the $\cldf$ $L$; therefore, some rates of the $\crdf$ $G$ must remain unused such that the effective rate becomes less than the $\cldf$.
	Similar to Theorem \ref{thm:lossless}, it follows from Lemma \ref{lmm:CRDFMaj}, that the best strategy is to remove the rates of the first blocks as the rates of the next blocks can compensate them.
	Therefore, the possible effective rate satisfying the security constraint is $G^{\rm eff}$.
	Regarding the distortion, since $D(\cdot)$ is a convex function, utilizing the majorization inequality (see Lemma \ref{lmm:MajIneq}), it can be shown that the best possible rate profile, with causality constraint, is $\hat{G}^{\rm eff}$, the concave-hull of the effective rate profile.

The next corollary states that, as expected, the result of Theorem \ref{thm:R(D)} for Hamming distortion, with distortion $0$, is the same as the result of Theorem \ref{thm:lossless}.
\begin{corollary} \label{crl:Lossless=Lossy for Hamming}
	An equivalent form of \eqref{eqn:F1-F0<H} in Theorem \ref{thm:lossless} is
	\begin{equation} \label{eqn:Geff1-0>H}
		G^{\rm eff}(1) - (1-\alpha)\H{X} 
		\geq G^{\rm eff}(\alpha),
		\qquad \forall \alpha\in[0,1],
	\end{equation}
	where $G^\mathrm{eff}$ is defined in \eqref{eqn:GL}.
	Further, \eqref{eqn:Geff1-0>H} follows from \eqref{eqn:intDG<d} in Theorem \ref{thm:R(D)} for Hamming distortion and $\bar{d} = 0$, where the Hamming distortion, $d(x,\hat{x})=0$ if $x=\hat{x}$ and $d(x,\hat{x})=1$ otherwise.
\end{corollary}

\begin{IEEEproof}
	See Section \ref{prf:crl:Lossless=Lossy for Hamming}.
\end{IEEEproof}

Based on Theorem \ref{thm:R(D)}, for a $\crdf$ $G$, only the concave hull of its effective rate, $\hat{G}^{\rm eff}$, is important. For instance, assume
\begin{equation*}
	G_1(\alpha) =
	\begin{cases}
		2\alpha & \alpha\in[0, 0.5),\\
		2 & \alpha\in[0.5, 1],
	\end{cases}
	\qquad
	G_2(\alpha) =
	\begin{cases}
		0 & \alpha\in[0, 0.5),\\
		2 & \alpha\in[0.5, 1],
	\end{cases}
	\qquad L(\alpha) =
	\begin{cases}
		5\alpha & \alpha\in[0, 0.2),\\
		1 & \alpha\in[0.2, 1].
	\end{cases}
\end{equation*}
Hence,
\begin{equation*}
	G_1^{\rm eff}(\alpha) 
	= G_2^{\rm eff}(\alpha) =
	\begin{cases}
		0 & \alpha\in[0, 0.5),\\
		1 & \alpha\in[0.5, 1].
	\end{cases}
\end{equation*}
Therefore, both of them give the same result.
So, in this case, due to the security constraint, increasing the rates does not help and they are redundant.
Another example is as follows.
Let
\begin{equation*}
	G_3(\alpha) =
	\begin{cases}
		4\alpha & \alpha\in[0, 0.5),\\
		2 & \alpha\in[0.5, 1].
	\end{cases}
	\Longrightarrow G_3^{\rm eff}(\alpha) =
	\begin{cases}
		0 & \alpha\in [0,0.25) \\
		4\alpha - 1 & \alpha\in[0, 0.5),\\
		1 & \alpha\in[0.5, 1].
	\end{cases}
\end{equation*}
However, we have that
\begin{equation*}
	\hat{G}_2^{\rm eff}(\alpha) 
	= \hat{G}_3^{\rm eff}(\alpha) =
	\begin{cases}
		2\alpha & \alpha\in[0, 0.5),\\
		1 & \alpha\in[0.5, 1].
	\end{cases}
\end{equation*}
Thus, they have the same achievability results despite having different effective rate.

In what follows, we give two examples to demonstrate the utility of Theorem \ref{thm:R(D)} and Corollary \ref{crl:Lossless=Lossy for Hamming}.

\begin{example}[Erasure distortion] \label{eg:erasure}
	Consider an $\IID$ sequence of $\rv$s $X_1,X_2,\ldots$ with $\mathrm{Bernoulli}(1/2)$ distribution. 
	The output support $\hat{\mathcal{X}}$ is $\{0,1,e\}$ and the distortion function is the \emph{erasure distortion} $d(x,\hat{x})$ as $d(0,0)=d(1,1)=0, d(0,1)=d(1,0)=\infty, d(0,e)=d(1,e)=1$.
	The rate-distortion function for this problem is \cite[Remark 3.9]{ElGamal:2011}
	\begin{equation} \label{eqn:R(D) erasure}
		R(D) = 
		\begin{cases}
			1-D & 0\leq D \leq 1, \\
			0 & D>1.
		\end{cases}
	\end{equation}
	From Lemma \ref{lmm:Linear R(D)}, a $\crdf$ $G$ is achievable, given $\cldf$ $L$, with distortion $\bar{d}$, if and only if
	\begin{equation*}
		G^\mathrm{eff}(1) - G^\mathrm{eff}(\alpha)
		\geq 1 - \bar{d} - \alpha,
		\qquad \forall\alpha\in[0,1-\bar{d}],
	\end{equation*}
	where $G^\mathrm{eff}$ is defined in \eqref{eqn:GL}.
	Intuitively, the result can be obtained from \eqref{eqn:R(D) erasure}; 
	to compress the source, with distortion $\bar{d}\leq 1$, we need to losslessly compress an arbitrary set of $(1-\bar{d})nk$ symbols of the source, and do not compress the other symbols.\\
	In our sequential framework, if $G^\mathrm{eff}$ is able to losslessly encode an arbitrary $(1-\bar{d})nk$ symbols of the source, it is also able to encode the first $(1-\bar{d})k$ blocks of the source because the rates from last blocks can be used to encode the first blocks, but not the other way.
	Hence, the optimal way is to use all possible rates to losslessly encode the first $(1-\bar{d})k$ blocks.\\
	Hence, from Corollary \ref{crl:Lossless=Lossy for Hamming}, we have 
	\begin{equation*}
		G^\mathrm{eff}(1) - G^\mathrm{eff}(\alpha)
		\geq (1 - \bar{d} - \alpha)\H{X}
		= 1 - \bar{d} - \alpha,
		\qquad \forall\alpha\in[0,1-\bar{d}],
	\end{equation*}
	where the first inequality follows from the fact that we use all $G^\mathrm{eff}(1)$ rates for the first $k(1-\bar{d})$ blocks; as a result, it is the same as having $G^\mathrm{eff}(1-\bar{d})=G^\mathrm{eff}(1)$.
	An example is illustrated in Fig. \ref{fig:eg:erasure} showing that only the effective $\crdf$s under the upper bound are achievable.
	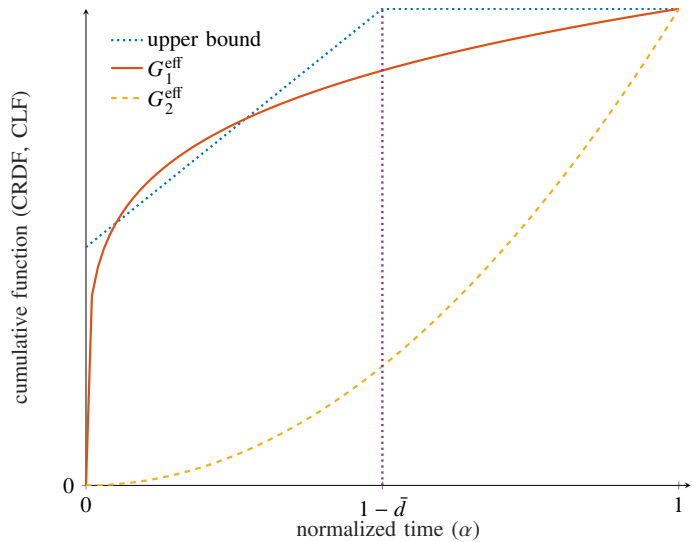
\begin{figure}
		\centering
		{
%
%
\definecolor{mycolor1}{rgb}{0.00000,0.44700,0.74100}%
\definecolor{mycolor2}{rgb}{0.85000,0.32500,0.09800}%
\definecolor{mycolor3}{rgb}{0.92900,0.69400,0.12500}%
\definecolor{mycolor4}{rgb}{0.49400,0.18400,0.55600}%
\begin{tikzpicture}[scale=0.7]

\begin{axis}[%
width=4.521in,
height=3.566in,
at={(0.758in,0.481in)},
scale only axis,
xmin=0,
xmax=1.02,
xtick={0,0.5,1},
xticklabels={{0},{$1-\bar{d}$},{1}},
xlabel style={font=\color{white!15!black}},
xlabel={normalized time ($\alpha$)},
ymin=0,
ymax=1,
ytick={0},
ylabel style={font=\color{white!15!black}},
ylabel={cumulative function ($\mathrm{CRDF}$, $\mathrm{CLF}$)},
axis x line=bottom,
axis y line=left,
legend style={at={(0.03,0.97)}, anchor=north west, legend cell align=left, align=left, fill=none, draw=none}
]
\addplot [color=mycolor1, dotted, line width=1.2pt]
  table[row sep=crcr]{%
0	0.5\\
0.01	0.51\\
0.02	0.52\\
0.03	0.53\\
0.04	0.54\\
0.05	0.55\\
0.06	0.56\\
0.07	0.57\\
0.08	0.58\\
0.09	0.59\\
0.1	0.6\\
0.11	0.61\\
0.12	0.62\\
0.13	0.63\\
0.14	0.64\\
0.15	0.65\\
0.16	0.66\\
0.17	0.67\\
0.18	0.68\\
0.19	0.69\\
0.2	0.7\\
0.21	0.71\\
0.22	0.72\\
0.23	0.73\\
0.24	0.74\\
0.25	0.75\\
0.26	0.76\\
0.27	0.77\\
0.28	0.78\\
0.29	0.79\\
0.3	0.8\\
0.31	0.81\\
0.32	0.82\\
0.33	0.83\\
0.34	0.84\\
0.35	0.85\\
0.36	0.86\\
0.37	0.87\\
0.38	0.88\\
0.39	0.89\\
0.4	0.9\\
0.41	0.91\\
0.42	0.92\\
0.43	0.93\\
0.44	0.94\\
0.45	0.95\\
0.46	0.96\\
0.47	0.97\\
0.48	0.98\\
0.49	0.99\\
0.5	1\\
0.51	1\\
0.52	1\\
0.53	1\\
0.54	1\\
0.55	1\\
0.56	1\\
0.57	1\\
0.58	1\\
0.59	1\\
0.6	1\\
0.61	1\\
0.62	1\\
0.63	1\\
0.64	1\\
0.65	1\\
0.66	1\\
0.67	1\\
0.68	1\\
0.69	1\\
0.7	1\\
0.71	1\\
0.72	1\\
0.73	1\\
0.74	1\\
0.75	1\\
0.76	1\\
0.77	1\\
0.78	1\\
0.79	1\\
0.8	1\\
0.81	1\\
0.82	1\\
0.83	1\\
0.84	1\\
0.85	1\\
0.86	1\\
0.87	1\\
0.88	1\\
0.89	1\\
0.9	1\\
0.91	1\\
0.92	1\\
0.93	1\\
0.94	1\\
0.95	1\\
0.96	1\\
0.97	1\\
0.98	1\\
0.99	1\\
1	1\\
};
\addlegendentry{upper bound}

\addplot [color=mycolor2, line width=1.2pt]
  table[row sep=crcr]{%
0	0\\
0.01	0.398107170553497\\
0.02	0.457305051927326\\
0.03	0.495934419641283\\
0.04	0.525305560880753\\
0.05	0.549280271653059\\
0.06	0.569679052028351\\
0.07	0.587515877741196\\
0.08	0.603417633654516\\
0.09	0.617800850567412\\
0.1	0.630957344480193\\
0.11	0.643100040646092\\
0.12	0.654389389941237\\
0.13	0.664949510209749\\
0.14	0.674878522295951\\
0.15	0.684255428918632\\
0.16	0.693144843155146\\
0.17	0.70160032942779\\
0.18	0.709666820762555\\
0.19	0.71738240420137\\
0.2	0.724779663677696\\
0.21	0.731886706417576\\
0.22	0.738727958788692\\
0.23	0.745324791387853\\
0.24	0.751696015753013\\
0.25	0.757858283255199\\
0.26	0.763826408534023\\
0.27	0.769613634072608\\
0.28	0.775231848384189\\
0.29	0.780691767293011\\
0.3	0.786003085596623\\
0.31	0.791174604765119\\
0.32	0.796214341106995\\
0.33	0.801129617900834\\
0.34	0.80592714427908\\
0.35	0.810613083098949\\
0.36	0.815193109605923\\
0.37	0.819672462357765\\
0.38	0.824055987609932\\
0.39	0.828348178150469\\
0.4	0.832553207401873\\
0.41	0.836674959469707\\
0.42	0.840717055706067\\
0.43	0.844682878264835\\
0.44	0.848575591050888\\
0.45	0.852398158403826\\
0.46	0.856153361805736\\
0.47	0.859843814860078\\
0.48	0.863471976753311\\
0.49	0.867040164381123\\
0.5	0.870550563296124\\
0.51	0.874005237612637\\
0.52	0.877406138986326\\
0.53	0.880755114771071\\
0.54	0.884053915442495\\
0.55	0.887304201366326\\
0.56	0.890507548980229\\
0.57	0.893665456449411\\
0.58	0.896779348849209\\
0.59	0.899850582921625\\
0.6	0.902880451447434\\
0.61	0.905870187270786\\
0.62	0.908820967009122\\
0.63	0.911733914477673\\
0.64	0.914610103854653\\
0.65	0.917450562610498\\
0.66	0.920256274222091\\
0.67	0.923028180690743\\
0.68	0.925767184880837\\
0.69	0.92847415269433\\
0.7	0.931149915094838\\
0.71	0.933795269993701\\
0.72	0.936410984009241\\
0.73	0.938997794109363\\
0.74	0.941556409146734\\
0.75	0.944087511294902\\
0.76	0.946591757392986\\
0.77	0.949069780205868\\
0.78	0.951522189606235\\
0.79	0.953949573684232\\
0.8	0.956352499790037\\
0.81	0.958731515514183\\
0.82	0.961087149610063\\
0.83	0.963419912862702\\
0.84	0.96573029890751\\
0.85	0.968018785002481\\
0.86	0.970285832756977\\
0.87	0.972531888820021\\
0.88	0.974757385530792\\
0.89	0.976962741533791\\
0.9	0.979148362360977\\
0.91	0.981314640982991\\
0.92	0.98346195833143\\
0.93	0.985590683793988\\
0.94	0.987701175684147\\
0.95	0.989793781686989\\
0.96	0.991868839282566\\
0.97	0.993926676148203\\
0.98	0.995967610540955\\
0.99	0.997991951661426\\
1	1\\
};
\addlegendentry{$G^\mathrm{eff}_1$}

\addplot [color=mycolor3, dashed, line width=1.2pt]
  table[row sep=crcr]{%
0	0\\
0.01	0.0001\\
0.02	0.0004\\
0.03	0.0009\\
0.04	0.0016\\
0.05	0.0025\\
0.06	0.0036\\
0.07	0.0049\\
0.08	0.0064\\
0.09	0.0081\\
0.1	0.01\\
0.11	0.0121\\
0.12	0.0144\\
0.13	0.0169\\
0.14	0.0196\\
0.15	0.0225\\
0.16	0.0256\\
0.17	0.0289\\
0.18	0.0324\\
0.19	0.0361\\
0.2	0.04\\
0.21	0.0441\\
0.22	0.0484\\
0.23	0.0529\\
0.24	0.0576\\
0.25	0.0625\\
0.26	0.0676\\
0.27	0.0729\\
0.28	0.0784\\
0.29	0.0841\\
0.3	0.09\\
0.31	0.0961\\
0.32	0.1024\\
0.33	0.1089\\
0.34	0.1156\\
0.35	0.1225\\
0.36	0.1296\\
0.37	0.1369\\
0.38	0.1444\\
0.39	0.1521\\
0.4	0.16\\
0.41	0.1681\\
0.42	0.1764\\
0.43	0.1849\\
0.44	0.1936\\
0.45	0.2025\\
0.46	0.2116\\
0.47	0.2209\\
0.48	0.2304\\
0.49	0.2401\\
0.5	0.25\\
0.51	0.2601\\
0.52	0.2704\\
0.53	0.2809\\
0.54	0.2916\\
0.55	0.3025\\
0.56	0.3136\\
0.57	0.3249\\
0.58	0.3364\\
0.59	0.3481\\
0.6	0.36\\
0.61	0.3721\\
0.62	0.3844\\
0.63	0.3969\\
0.64	0.4096\\
0.65	0.4225\\
0.66	0.4356\\
0.67	0.4489\\
0.68	0.4624\\
0.69	0.4761\\
0.7	0.49\\
0.71	0.5041\\
0.72	0.5184\\
0.73	0.5329\\
0.74	0.5476\\
0.75	0.5625\\
0.76	0.5776\\
0.77	0.5929\\
0.78	0.6084\\
0.79	0.6241\\
0.8	0.64\\
0.81	0.6561\\
0.82	0.6724\\
0.83	0.6889\\
0.84	0.7056\\
0.85	0.7225\\
0.86	0.7396\\
0.87	0.7569\\
0.88	0.7744\\
0.89	0.7921\\
0.9	0.81\\
0.91	0.8281\\
0.92	0.8464\\
0.93	0.8649\\
0.94	0.8836\\
0.95	0.9025\\
0.96	0.9216\\
0.97	0.9409\\
0.98	0.9604\\
0.99	0.9801\\
1	1\\
};
\addlegendentry{$G^\mathrm{eff}_2$}

\addplot [color=mycolor4, dotted, line width=1.2pt, forget plot]
  table[row sep=crcr]{%
0.5	0\\
0.5	0.0101010101010101\\
0.5	0.0202020202020202\\
0.5	0.0303030303030303\\
0.5	0.0404040404040404\\
0.5	0.0505050505050505\\
0.5	0.0606060606060606\\
0.5	0.0707070707070707\\
0.5	0.0808080808080808\\
0.5	0.0909090909090909\\
0.5	0.101010101010101\\
0.5	0.111111111111111\\
0.5	0.121212121212121\\
0.5	0.131313131313131\\
0.5	0.141414141414141\\
0.5	0.151515151515152\\
0.5	0.161616161616162\\
0.5	0.171717171717172\\
0.5	0.181818181818182\\
0.5	0.191919191919192\\
0.5	0.202020202020202\\
0.5	0.212121212121212\\
0.5	0.222222222222222\\
0.5	0.232323232323232\\
0.5	0.242424242424242\\
0.5	0.252525252525253\\
0.5	0.262626262626263\\
0.5	0.272727272727273\\
0.5	0.282828282828283\\
0.5	0.292929292929293\\
0.5	0.303030303030303\\
0.5	0.313131313131313\\
0.5	0.323232323232323\\
0.5	0.333333333333333\\
0.5	0.343434343434343\\
0.5	0.353535353535354\\
0.5	0.363636363636364\\
0.5	0.373737373737374\\
0.5	0.383838383838384\\
0.5	0.393939393939394\\
0.5	0.404040404040404\\
0.5	0.414141414141414\\
0.5	0.424242424242424\\
0.5	0.434343434343434\\
0.5	0.444444444444444\\
0.5	0.454545454545455\\
0.5	0.464646464646465\\
0.5	0.474747474747475\\
0.5	0.484848484848485\\
0.5	0.494949494949495\\
0.5	0.505050505050505\\
0.5	0.515151515151515\\
0.5	0.525252525252525\\
0.5	0.535353535353535\\
0.5	0.545454545454545\\
0.5	0.555555555555556\\
0.5	0.565656565656566\\
0.5	0.575757575757576\\
0.5	0.585858585858586\\
0.5	0.595959595959596\\
0.5	0.606060606060606\\
0.5	0.616161616161616\\
0.5	0.626262626262626\\
0.5	0.636363636363636\\
0.5	0.646464646464647\\
0.5	0.656565656565657\\
0.5	0.666666666666667\\
0.5	0.676767676767677\\
0.5	0.686868686868687\\
0.5	0.696969696969697\\
0.5	0.707070707070707\\
0.5	0.717171717171717\\
0.5	0.727272727272727\\
0.5	0.737373737373737\\
0.5	0.747474747474748\\
0.5	0.757575757575758\\
0.5	0.767676767676768\\
0.5	0.777777777777778\\
0.5	0.787878787878788\\
0.5	0.797979797979798\\
0.5	0.808080808080808\\
0.5	0.818181818181818\\
0.5	0.828282828282828\\
0.5	0.838383838383838\\
0.5	0.848484848484849\\
0.5	0.858585858585859\\
0.5	0.868686868686869\\
0.5	0.878787878787879\\
0.5	0.888888888888889\\
0.5	0.898989898989899\\
0.5	0.909090909090909\\
0.5	0.919191919191919\\
0.5	0.929292929292929\\
0.5	0.939393939393939\\
0.5	0.94949494949495\\
0.5	0.95959595959596\\
0.5	0.96969696969697\\
0.5	0.97979797979798\\
0.5	0.98989898989899\\
0.5	1\\
};
\end{axis}
\end{tikzpicture}%
}
		\caption{An illustration of Example \ref{eg:erasure}. 
		The upper bound is $\min\{\alpha + G^\mathrm{eff}(1)-1+\bar{d}, G^\mathrm{eff}(1)\}$.
		$\crdf$ $G_1^\mathrm{eff}$ is not achievable while $\crdf$ $G_2^\mathrm{eff}$ is achievable.}
		\label{fig:eg:erasure}
	\end{figure}
\end{example}

\begin{example}[Log-loss distortion] \label{eg:logloss}
	Consider a sequence of $\IID$ $\rv$s $X_1,X_2,\ldots$ with $\pmf$ $p(x)$ over the finite domain $\mathcal{X}$.
	The output support is the set of all possible $\pmf$s over $\mathcal{X}$.
	The distortion $d(x,\hat{p})$ is $-\log p(x)$ for any $\pmf$ $\hat{p}$ defined over over $\mathcal{X}$.
	The rate-distortion function for this problem is \cite{Courtade13, Shkel17}
	\begin{equation} \label{eqn:R(D) logloss}
		R(D) = 
		\begin{cases}
			\H{X}-D & 0\leq D \leq \H{X}, \\
			0 & D>\H{X},
		\end{cases}
	\end{equation}
	where $X$ is a $\rv$ with $\pmf$ $p(x)$.
	From Lemma \ref{lmm:Linear R(D)}, we have that a $\crdf$ $G$ is achievable, given $\cldf$ $L$, with distortion $\bar{d}$, if and only if
	\begin{equation*}
		G^\mathrm{eff}(1) - G^\mathrm{eff}(\alpha)
		\geq (1 - \alpha) \H{X} - \bar{d},
		\qquad \forall\alpha\in\left[0,1 - \frac{\bar{d}}{\H{X}}\right],
	\end{equation*}
	where $G^\mathrm{eff}$ is defined in \eqref{eqn:GL}.
	Intuitively, the result can be obtained from \eqref{eqn:R(D) logloss};
	to compress the source, with distortion $\bar{d}\leq \H{X}$, we need to send $\H{X} - \bar{d}$ bits per symbol of the source.\\
	To do so, as we explained in the previous example, since the rates from last blocks can be used to encode the first blocks, but not the other way around, the optimal approach is to losslessly compress the first $k(\H{X} - \bar{d})/\H{X}$ blocks of the source.
	Hence, from Corollary \ref{crl:Lossless=Lossy for Hamming}, we have 
	\begin{equation*}
		G^\mathrm{eff}(1) - G^\mathrm{eff}(\alpha)
		\geq \left( \frac{\H{X} - \bar{d}}{\H{X}} - \alpha \right) \H{X}
		= (1 - \alpha) \H{X} - \bar{d},
		\qquad \forall\alpha\in\left[0,1 - \frac{\bar{d}}{\H{X}}\right],
	\end{equation*}
	where the first inequality follows from the fact that we use all $G^\mathrm{eff}(1)$ rates for the first $k(\H{X} - \bar{d})/\H{X}$ blocks; as a result, it is the same as having $G^\mathrm{eff}(1-\bar{d}/\H{X})=G^\mathrm{eff}(1)$.
	As an example, consider $\cldf$ $L(\alpha)$ which is $0$ for $\alpha\leq\alpha'$ and large enough for $\alpha>\alpha'$.
	Hence, any effective $G^\mathrm{eff}(\alpha)$ is $0$ for $\alpha\leq\alpha'$.
	As illustrated in Fig. \ref{fig:eg:logloss}, if $1-\bar{d}/\H{X}\leq\alpha'$, then effective $\crdf$ $G^\mathrm{eff}$ is achievable if and only if $G^\mathrm{eff}(1) \geq \H{X} - \bar{d}$ because the upper bound becomes negative for $\alpha=0$ otherwise.
	\begin{figure}
		\centering
		{\input{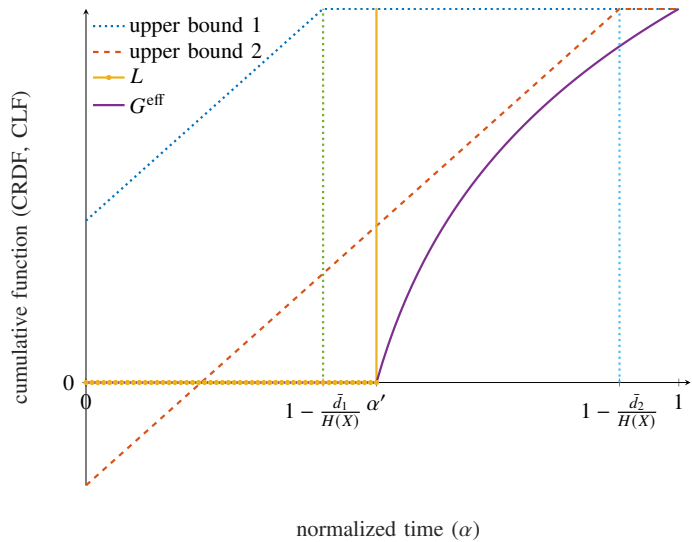}}
		\caption{An illustration of Example \ref{eg:logloss}.
		The upper bounds are $\min\{\alpha\H{X} + G^\mathrm{eff}(1) - \H{X} + \bar{d}, G^\mathrm{eff}(1)\}$, for $\bar{d}=\bar{d}_1$ and $\bar{d}=\bar{d}_2$, respectively.
		For the given $\cldf$ $L$, $\crdf$ $G^\mathrm{eff}$, is achievable with distortion $\bar{d}_1$ while it is not achievable with distortion $\bar{d}_2$.}
		\label{fig:eg:logloss}
	\end{figure}
\end{example}

\section{Proofs} \label{sec:proofs}
In this section, we prove the main results of the paper.
\subsection{Proof of Theorem \ref{thm:lossless}} \label{subsec:prf:thm:lossless}
First we prove the converse part, which claims that for any $\crdf$ $G$ achievable with $\cldf$ $L$, \eqref{eqn:F1-F0<H} must be satisfied.
Then, we prove the achievability, which claims that if a $\crdf$ $G$ satisfies \eqref{eqn:F1-F0<H}, given $\cldf$ $L$, then it is achievable.

\textbf{Converse}: For all $n,k\in\mathbb{N}$ and $j\in\{1,\ldots,k-1\}$, we have that
\begin{IEEEeqnarray}{rCl}
	n k \sum_{i=j+1}^k{R_i}
	&\geq& \sum_{i=j+1}^k{\H{M_i}}
	\geq \Hb{M_{j+1}^k}
	\geq \Hb{M_{j+1}^k \vert M^j}
	= \Hb{X^{n k}, M_{j+1}^k \vert M^j}
	- \Hb{X^{n k} \vert M^k} \nonumber\\
	&\geq& \Hb{X^{n k}, M_{j+1}^k \vert M^j}
	- nk\epsilon_n \label{eqn:Fano}\\
	&\geq& \Hb{X^{n k} \vert M^j}
	- nk\epsilon_n
	= \Hb{X^{n j} \vert M^j}
	+ \Hb{X_{n j + 1}^{n k} \vert M^j, X^{n j}}
	- nk\epsilon_n \nonumber\\
	&\geq& \max\left\{0, \Hb{X^{n j}} - n k L\left(\tfrac{j}{k}\right) \right\}
	+ \Hb{X_{n j + 1}^{n k} \vert M^j, X^{n j}}
	- nk\epsilon_n \label{eqn:H(Xnj|M)<H-L} \\
	&=&  \max\left\{0, \Hb{X^{n j}} - n k L\left(\tfrac{j}{k}\right) \right\}
	+ \Hb{X_{n j + 1}^{n k}}
	- nk\epsilon_n \label{eqn:HX|M=HX} \\
	&=&  \max\left\{0, n j \H{X} - n k L\left(\tfrac{j}{k}\right) \right\}
	+ n(k-j) \H{X}
	- nk\epsilon_n, \label{eqn:H(Xnj),H(Xnj^nk) iid}
\end{IEEEeqnarray}
where \eqref{eqn:Fano} follows from Fano's inequality \cite[Theorem 2.10.1]{Cover06} with
\begin{equation} \label{eqn:en->0 lossless}
	\epsilon_{n}
	= \Pr\{X^{nk} \neq \hat{X}^{nk}\}\log\lvert\mathcal{X}\rvert + \frac{1}{nk}
	\Longrightarrow \lim_{n\to\infty}\epsilon_n = 0,
	\quad \forall k\in\mathbb{N},
\end{equation}
which follows from Definition \ref{def:Achievable Lossless Codes} because $\Pr\{X^{nk} \neq \hat{X}^{nk}\}$ vanishes as $n\to\infty$;
\eqref{eqn:H(Xnj|M)<H-L} follows from the definition of $L$ in \eqref{eqn:I<L lossless};
\eqref{eqn:HX|M=HX} is true because $M^j$ is a function of $X^{n j}$, according to Definition \ref{def:Code}, and $X^{n j}$ is independent of $X_{n j+1}^{n k}$;
and \eqref{eqn:H(Xnj),H(Xnj^nk) iid} follows because $X^{nk}$ is an $\IID$ sequence.
Therefore, from the definition of $G$ in \eqref{eqn:def R} and using \eqref{eqn:en->0 lossless}, as $n\to\infty$, we obtain that, for $j=0,\ldots,k-1$,
\begin{equation} \label{eqn:F1-F0<H Q}
	G(1) - G\left(\tfrac{j}{k}\right)
	\geq \max\left\{0, \frac{j}{k} \H{X} - L\left(\tfrac{j}{k}\right) \right\}
	+ \left(1-\frac{j}{k}\right) \H{X}.
\end{equation}
To show \eqref{eqn:F1-F0<H} for $\alpha\in[0,1)$, assume that \eqref{eqn:F1-F0<H} is violated for some $\alpha\in[0,1)$.
Therefore, 
\begin{equation*}
	G(1) - G(\alpha)
	- \max\{(1-\alpha) \H{X}, \H{X} - L(\alpha) \}
	< 0
	\Rightarrow
	\begin{cases}
		G(1) - G(\alpha) - (1-\alpha) \H{X} < 0, \;\text{or} \\
		G(1) - G(\alpha) - \H{X} + L(\alpha) < 0.
	\end{cases}
\end{equation*}
Both functions $G$ and $L$ are continuous from the right because of their regularity condition (see Definitions \ref{def:CRDF} and \ref{def:CLDF});
as a result, $G(1) - G(\alpha) - (1-\alpha) \H{X}$ and $G(1) - G(\alpha) - \H{X} + L(\alpha)$ are continuous from the righ.
Hence, in each case that \eqref{eqn:F1-F0<H} is violated, there exists some $j,k\in\mathbb{N}$ such that \eqref{eqn:F1-F0<H Q} is also violated, which is a contradiction. \\
For $\alpha = 1$, it is sufficient to show $L(1) \geq \H{X}$, which follows from
\begin{equation*}
	nk L(1)
	\stackrel{(a)}{\geq} \Ib{X^{nk}}{M^k}
	= nk\Hb{X} - \Hb{X^{nk} \vert M^k}
	\stackrel{(b)}{\geq} n k \H{X} - n k \epsilon_n,
\end{equation*}
where $(a)$ follows from the definition of $L$ in \eqref{eqn:I<L lossless}
and $(b)$ follows the same as \eqref{eqn:Fano}.
Hence, the result follows as $n\to\infty$ utilizing \eqref{eqn:en->0 lossless}.
Thus, \eqref{eqn:F1-F0<H} is proved.

\textbf{Achievability}:
From \eqref{eqn:F1-F0<H}, for $\alpha=0$, we obtain that $G(1) \geq \H{X}$.
Hence, utilizing Lemma \ref{lmm:CRDFshift}, it suffices to prove that the $\bar{G}^{\rm eff}$ is achievable, where $\bar{G}^{\rm eff}$ is defined in \eqref{eqn:barGeff}.
Hence, from \eqref{eqn:F1-F0<H}, we can write, for all $\alpha\in[0,1]$, \\
\begin{IEEEeqnarray}{l}
	\bar{G}^{\rm eff}(\alpha) \leq \alpha \H{X}, \label{eqn:Geff<aH} \\
	\bar{G}^{\rm eff}(\alpha) \leq L(\alpha). \label{eqn:Geff<L lossless}
\end{IEEEeqnarray}
Define $\crdf$ $\tilde{G}\colon \alpha\mapsto\alpha\H{X}$.
From Lemma \ref{lmm:CRDFMaj} and \eqref{eqn:Geff<aH}, we obtain that if $\tilde{G}$ satisfies \eqref{eqn:Pe->0}, then there exists a sequence of coding schemes satisfying \eqref{eqn:Pe->0} with $\crdf$ $\bar{G}^{\rm eff}$.
To show that, for any $k\in\mathbb{N}$, $\tilde{G}$ satisfies \eqref{eqn:Pe->0}, we have from Definition \ref{def:Code}, $\tilde{R}_i:=\tilde{G}(i/k) - \tilde{G}((i-1)/k) = \H{X}/k$.
Hence, from the classical lossless source coding theorem \cite[Theorem 3.4]{ElGamal:2011}, we obtain that for all $i\in\{1,\ldots,k\}$, there exists a sequence of encoders and decoders to compress $X_{(i-1)n+1}^{i n}$ with rate $k \tilde{R}_i$ and vanishing probability of error $\Pr\big\{ X_{(i-1)n+1}^{i n} \neq \hat{X}_{(i-1)n+1}^{i n}\big\}$ as $n\to\infty$.
Hence, \eqref{eqn:Pe->0} follows from the union bound for all $k$ blocks. \\
Now, it only remains to show that $\bar{G}^{\rm eff}$ satisfies the leakage constraint \eqref{eqn:I<L lossless}.
To this end, we have that for all $k\in\mathbb{N}$
\begin{equation*}
	\Ib{X^{j k}}{M^j}
	\leq \Hb{M^j}
	\leq nk\sum_{i=1}^j R_i
	= nk\bar{G}^{\rm eff}\left(\tfrac{j}{k}\right)
	\leq nk L\left(\tfrac{j}{k}\right),
	\qquad \forall j\in\{1,\ldots,k\},
\end{equation*}
where the last inequality follows from \eqref{eqn:Geff<L lossless}.
Thus, the achievability part is proved.
\qed

\subsection{Proof of Theorem \ref{thm:R(D)}} \label{subsec:prf:R(D)}
First we prove the converse part, which claims that if $\crdf$ $G$ is achievable with $\cldf$ $L$, then for any $k\in\mathbb{N}$, the sequence of $(G,k,n)$-source codes satisfying the distortion constraint must satisfy \eqref{eqn:intDG<d}.
After that, we prove the achievability part, which claims that if a $\crdf$ $G$ satisfies \eqref{eqn:intDG<d} for a given $\cldf$ $L$, then it is achievable.
For the distortion, we use the following notation:
\begin{equation} \label{eqn:partiald}
	d(x_a^b, \hat{x}_a^b)
	:= \frac{1}{b-a+1}\sum_{i=a}^b{d(x_i,\hat{x}_i)}.
\end{equation}

\begin{figure}
	\begin{subfigure}{0.5 \textwidth}
		\centering
%
%
\definecolor{mycolor1}{rgb}{0.00000,0.44700,0.74100}%
\definecolor{mycolor2}{rgb}{0.85000,0.32500,0.09800}%
\definecolor{mycolor3}{rgb}{0.92900,0.69400,0.12500}%
\begin{tikzpicture}[scale=0.5]

\begin{axis}[%
width=4.521in,
height=3.566in,
at={(0.758in,0.481in)},
scale only axis,
xmin=0,
xmax=1.02,
xlabel style={font=\color{white!15!black}},
xlabel={$\alpha$},
ymin=0,
ymax=2.249998,
ytick={0},
yticklabels={{0}},
ylabel style={font=\color{white!15!black}},
ylabel={cumulative function},
axis x line=bottom,
axis y line=left,
legend style={at={(0.03,0.97)}, anchor=north west, legend cell align=left, align=left, fill=none, draw=none}
]
\addplot [color=mycolor1, line width=1.2pt]
  table[row sep=crcr]{%
0	0\\
0.01	0\\
0.02	0\\
0.03	0\\
0.04	0\\
0.05	0\\
0.06	0\\
0.07	0.0370008618806215\\
0.08	0.0911908618806215\\
0.09	0.142860861880621\\
0.1	0.192070861880621\\
0.11	0.238880861880621\\
0.12	0.283350861880621\\
0.13	0.325540861880621\\
0.14	0.365510861880621\\
0.15	0.403320861880621\\
0.16	0.439030861880621\\
0.17	0.472700861880621\\
0.18	0.504390861880621\\
0.19	0.534160861880621\\
0.2	0.562070861880621\\
0.21	0.588180861880621\\
0.22	0.612550861880621\\
0.23	0.635240861880621\\
0.24	0.656310861880621\\
0.25	0.675820861880621\\
0.26	0.693830861880621\\
0.27	0.710400861880621\\
0.28	0.725590861880621\\
0.29	0.739460861880621\\
0.3	0.752070861880621\\
0.31	0.763480861880621\\
0.32	0.773750861880621\\
0.33	0.782940861880622\\
0.34	0.791110861880621\\
0.35	0.798320861880621\\
0.36	0.804630861880621\\
0.37	0.810100861880621\\
0.38	0.814790861880621\\
0.39	0.818760861880621\\
0.4	0.822070861880621\\
0.41	0.824780861880621\\
0.42	0.826950861880621\\
0.43	0.828640861880621\\
0.44	0.829910861880621\\
0.45	0.830820861880621\\
0.46	0.831430861880621\\
0.47	0.831800861880621\\
0.48	0.831990861880622\\
0.49	0.832060861880621\\
0.5	0.832068861880621\\
0.51	0.832076861880621\\
0.52	0.832132861880621\\
0.53	0.832284861880621\\
0.54	0.832580861880622\\
0.55	0.833068861880621\\
0.56	0.833796861880621\\
0.57	0.834812861880621\\
0.58	0.836164861880622\\
0.59	0.837900861880621\\
0.6	0.840068861880621\\
0.61	0.842716861880621\\
0.62	0.845892861880621\\
0.63	0.849644861880621\\
0.64	0.854020861880621\\
0.65	0.859068861880621\\
0.66	0.864836861880621\\
0.67	0.871372861880621\\
0.68	0.878724861880622\\
0.69	0.886940861880621\\
0.7	0.896068861880621\\
0.71	0.906156861880621\\
0.72	0.917252861880621\\
0.73	0.929404861880621\\
0.74	0.942660861880622\\
0.75	0.957068861880621\\
0.76	0.972676861880621\\
0.77	0.989532861880621\\
0.78	1.00768486188062\\
0.79	1.02718086188062\\
0.8	1.04806886188062\\
0.81	1.07039686188062\\
0.82	1.09421286188062\\
0.83	1.11956486188062\\
0.84	1.14650086188062\\
0.85	1.17506886188062\\
0.86	1.20531686188062\\
0.87	1.23729286188062\\
0.88	1.27104486188062\\
0.89	1.30662086188062\\
0.9	1.34406886188062\\
0.91	1.38343686188062\\
0.92	1.42477286188062\\
0.93	1.46812486188062\\
0.94	1.51354086188062\\
0.95	1.56106886188062\\
0.96	1.61075686188062\\
0.97	1.66265286188062\\
0.98	1.71680486188062\\
0.99	1.77326086188062\\
1	1.83206886188062\\
};
\addlegendentry{$G^{\rm eff}_k$}

\addplot[ycomb, color=mycolor2, line width=1.2pt, mark=x, mark options={solid, mycolor2}] table[row sep=crcr] {%
0.1	0.565706886188063\\
0.2	0.95417733101834\\
0.3	1.2\\
0.4	1.33776355864221\\
0.5	1.40205667245415\\
0.6	1.48805911033945\\
0.7	1.57406154822474\\
0.8	1.66006398611003\\
0.9	1.74606642399533\\
1	1.83206886188062\\
};
\addplot[forget plot, color=white!15!black, line width=1.2pt] table[row sep=crcr] {%
0	0\\
1	0\\
};
\addlegendentry{$\sum_{i=1}^j\tilde{R_i}^{\prime}$}

\addplot[ycomb, color=mycolor3, line width=1.2pt, mark=o, mark options={solid, mycolor3}] table[row sep=crcr] {%
0.1	0.565706886188063\\
0.2	0.95417733101834\\
0.3	1.2\\
0.4	1.33776355864221\\
0.5	1.40205667245415\\
0.6	1.42746800694498\\
0.7	1.44858622762388\\
0.8	1.5\\
0.9	1.61629798958253\\
1	1.83206886188062\\
};
\addplot[forget plot, color=white!15!black, line width=1.2pt] table[row sep=crcr] {%
0	0\\
1	0\\
};
\addlegendentry{$\sum_{i=1}^j\tilde{R_i}$}

\end{axis}
\end{tikzpicture}%
		\caption{Step \ref{itm:decR}}
	\end{subfigure}
	\hfill
	\begin{subfigure}{0.5 \textwidth}
		\centering
%
%
\definecolor{mycolor1}{rgb}{0.00000,0.44700,0.74100}%
\definecolor{mycolor2}{rgb}{0.85000,0.32500,0.09800}%
\definecolor{mycolor3}{rgb}{0.92900,0.69400,0.12500}%
\begin{tikzpicture}[scale=0.5]

\begin{axis}[%
width=4.521in,
height=3.566in,
at={(0.758in,0.481in)},
scale only axis,
xmin=0,
xmax=1.02,
xlabel style={font=\color{white!15!black}},
xlabel={$\alpha$},
ymin=0,
ymax=1.83206886188062,
ytick={0},
yticklabels={{0}},
ylabel style={font=\color{white!15!black}},
ylabel={cumulative function},
axis x line=bottom,
axis y line=left,
legend style={at={(0.03,0.97)}, anchor=north west, legend cell align=left, align=left, fill=none, draw=none}
]
\addplot [color=mycolor1, line width=1.2pt]
  table[row sep=crcr]{%
0	0\\
0.1	0\\
0.1	0.192070861880621\\
0.2	0.192070861880621\\
0.2	0.562070861880621\\
0.3	0.562070861880621\\
0.3	0.752070861880621\\
0.4	0.752070861880621\\
0.4	0.822070861880621\\
0.5	0.822070861880621\\
0.5	0.832068861880621\\
0.6	0.832068861880621\\
0.6	0.840068861880621\\
0.7	0.840068861880621\\
0.7	0.896068861880621\\
0.8	0.896068861880621\\
0.8	1.04806886188062\\
0.9	1.04806886188062\\
0.9	1.34406886188062\\
1	1.34406886188062\\
1	1.83206886188062\\
};
\addlegendentry{$\bar{G}^{\rm eff}_k$}

\addplot [color=mycolor2, line width=1.2pt]
  table[row sep=crcr]{%
0	0\\
0.01	0\\
0.02	0\\
0.03	0\\
0.04	0\\
0.05	0\\
0.06	0\\
0.07	0.0370008618806215\\
0.08	0.0911908618806215\\
0.09	0.142860861880621\\
0.1	0.192070861880621\\
0.11	0.238880861880621\\
0.12	0.283350861880621\\
0.13	0.325540861880621\\
0.14	0.365510861880621\\
0.15	0.403320861880621\\
0.16	0.439030861880621\\
0.17	0.472700861880621\\
0.18	0.504390861880621\\
0.19	0.534160861880621\\
0.2	0.562070861880621\\
0.21	0.588180861880621\\
0.22	0.612550861880621\\
0.23	0.635240861880621\\
0.24	0.656310861880621\\
0.25	0.675820861880621\\
0.26	0.693830861880621\\
0.27	0.710400861880621\\
0.28	0.725590861880621\\
0.29	0.739460861880621\\
0.3	0.752070861880621\\
0.31	0.763480861880621\\
0.32	0.773750861880621\\
0.33	0.782940861880622\\
0.34	0.791110861880621\\
0.35	0.798320861880621\\
0.36	0.804630861880621\\
0.37	0.810100861880621\\
0.38	0.814790861880621\\
0.39	0.818760861880621\\
0.4	0.822070861880621\\
0.41	0.824780861880621\\
0.42	0.826950861880621\\
0.43	0.828640861880621\\
0.44	0.829910861880621\\
0.45	0.830820861880621\\
0.46	0.831430861880621\\
0.47	0.831800861880621\\
0.48	0.831990861880622\\
0.49	0.832060861880621\\
0.5	0.832068861880621\\
0.51	0.832076861880621\\
0.52	0.832132861880621\\
0.53	0.832284861880621\\
0.54	0.832580861880622\\
0.55	0.833068861880621\\
0.56	0.833796861880621\\
0.57	0.834812861880621\\
0.58	0.836164861880622\\
0.59	0.837900861880621\\
0.6	0.840068861880621\\
0.61	0.842716861880621\\
0.62	0.845892861880621\\
0.63	0.849644861880621\\
0.64	0.854020861880621\\
0.65	0.859068861880621\\
0.66	0.864836861880621\\
0.67	0.871372861880621\\
0.68	0.878724861880622\\
0.69	0.886940861880621\\
0.7	0.896068861880621\\
0.71	0.906156861880621\\
0.72	0.917252861880621\\
0.73	0.929404861880621\\
0.74	0.942660861880622\\
0.75	0.957068861880621\\
0.76	0.972676861880621\\
0.77	0.989532861880621\\
0.78	1.00768486188062\\
0.79	1.02718086188062\\
0.8	1.04806886188062\\
0.81	1.07039686188062\\
0.82	1.09421286188062\\
0.83	1.11956486188062\\
0.84	1.14650086188062\\
0.85	1.17506886188062\\
0.86	1.20531686188062\\
0.87	1.23729286188062\\
0.88	1.27104486188062\\
0.89	1.30662086188062\\
0.9	1.34406886188062\\
0.91	1.38343686188062\\
0.92	1.42477286188062\\
0.93	1.46812486188062\\
0.94	1.51354086188062\\
0.95	1.56106886188062\\
0.96	1.61075686188062\\
0.97	1.66265286188062\\
0.98	1.71680486188062\\
0.99	1.77326086188062\\
1	1.83206886188062\\
};
\addlegendentry{$G^{\rm eff}_k$}

\addplot [color=mycolor3, line width=1.2pt]
  table[row sep=crcr]{%
0	0\\
0.01	0.0281137295726643\\
0.02	0.0562274591453286\\
0.03	0.0843411887179928\\
0.04	0.112454918290657\\
0.05	0.140568647863321\\
0.06	0.168682377435986\\
0.07	0.19679610700865\\
0.08	0.224909836581314\\
0.09	0.253023566153979\\
0.1	0.281137295726643\\
0.11	0.309251025299307\\
0.12	0.337364754871971\\
0.13	0.365478484444636\\
0.14	0.3935922140173\\
0.15	0.421705943589964\\
0.16	0.449819673162629\\
0.17	0.477933402735293\\
0.18	0.506047132307957\\
0.19	0.534160861880621\\
0.2	0.562070861880621\\
0.21	0.581780861880621\\
0.22	0.601490861880621\\
0.23	0.621200861880621\\
0.24	0.640910861880621\\
0.25	0.660620861880621\\
0.26	0.680330861880621\\
0.27	0.700040861880621\\
0.28	0.719750861880621\\
0.29	0.739460861880621\\
0.3	0.752070861880621\\
0.31	0.767499404737764\\
0.32	0.782927947594907\\
0.33	0.79835649045205\\
0.34	0.813785033309193\\
0.35	0.829213576166336\\
0.36	0.844642119023479\\
0.37	0.860070661880621\\
0.38	0.875499204737764\\
0.39	0.890927747594907\\
0.4	0.90635629045205\\
0.41	0.921784833309193\\
0.42	0.937213376166336\\
0.43	0.952641919023479\\
0.44	0.968070461880621\\
0.45	0.983499004737764\\
0.46	0.998927547594907\\
0.47	1.01435609045205\\
0.48	1.02978463330919\\
0.49	1.04521317616634\\
0.5	1.06064171902348\\
0.51	1.07607026188062\\
0.52	1.09149880473776\\
0.53	1.10692734759491\\
0.54	1.12235589045205\\
0.55	1.13778443330919\\
0.56	1.15321297616634\\
0.57	1.16864151902348\\
0.58	1.18407006188062\\
0.59	1.19949860473776\\
0.6	1.21492714759491\\
0.61	1.23035569045205\\
0.62	1.24578423330919\\
0.63	1.26121277616634\\
0.64	1.27664131902348\\
0.65	1.29206986188062\\
0.66	1.30749840473776\\
0.67	1.32292694759491\\
0.68	1.33835549045205\\
0.69	1.35378403330919\\
0.7	1.36921257616634\\
0.71	1.38464111902348\\
0.72	1.40006966188062\\
0.73	1.41549820473776\\
0.74	1.43092674759491\\
0.75	1.44635529045205\\
0.76	1.46178383330919\\
0.77	1.47721237616634\\
0.78	1.49264091902348\\
0.79	1.50806946188062\\
0.8	1.52349800473776\\
0.81	1.53892654759491\\
0.82	1.55435509045205\\
0.83	1.56978363330919\\
0.84	1.58521217616634\\
0.85	1.60064071902348\\
0.86	1.61606926188062\\
0.87	1.63149780473776\\
0.88	1.64692634759491\\
0.89	1.66235489045205\\
0.9	1.67778343330919\\
0.91	1.69321197616634\\
0.92	1.70864051902348\\
0.93	1.72406906188062\\
0.94	1.73949760473776\\
0.95	1.75492614759491\\
0.96	1.77035469045205\\
0.97	1.78578323330919\\
0.98	1.80121177616634\\
0.99	1.81664031902348\\
1	1.83206886188062\\
};
\addlegendentry{$\hat{G}^{\rm eff}_k$}

\end{axis}
\end{tikzpicture}%
		\caption{Step \ref{itm:existConcG}}
	\end{subfigure}
	\caption{An example illustrating a few steps of the proof of the converse part of Theorem \ref{thm:R(D)}.}
	\label{fig:ConvStepR(D)}
\end{figure}
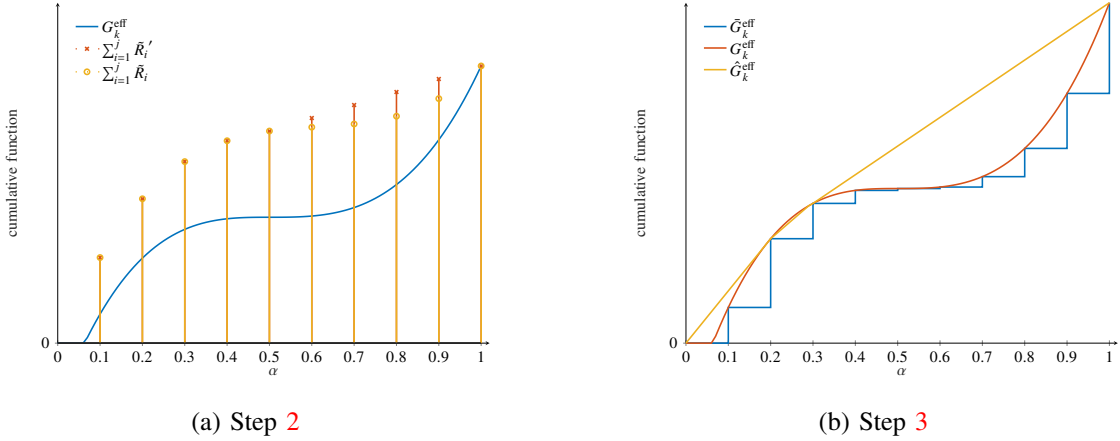

\textbf{Converse}:
The proof is divided in the following four steps.
For an illustration, see Fig. \ref{fig:ConvStepR(D)}.

\begin{enumerate}	
	\item\label{itm:ClasConv}
	In this step, we find a relation between the sequence $\tilde{R}_i$, which characterizes the effective rate per block, i.e., the minimum rate needed to satisfy the distortion constraint of the corresponding block, and the sequence $R_i$, which was defined in \eqref{eqn:def R}.
	We do not need to use all $\tilde{R}_i$ at the end of block $i$ and we can compress only a part of it at the end of the block and transfer the rates of the remaining part to the next blocks.
	This strategy helps to satisfy the leakage constraint.
	As a result, the sequence $R_i$ includes both a part $\tilde{R}_i$, and the rates transfered from the previous blocks.\\
	Formally, for any $k\in\mathbb{N}$, there exists a sequence $\tilde{R}_i$, for $i=1,\ldots,k$, such that
	\begin{IEEEeqnarray}{l}
		\sum_{i=j+1}^k{R_i}
		\geq \sum_{i=1}^k{\tilde{R}_i}
		- \min \left\{L\left(\frac{j}{k}\right), \sum_{i=1}^j{\tilde{R}_i}\right\},
		\quad j=1,\ldots,k-1, \label{eqn:G-Gj/k>sumRhat} \\
		\sum_{i=1}^k{\tilde{R}_i} \leq \min\left\{L(1), \sum_{i=1}^k{R_i}\right\}, \label{eqn:sumRt<L1,sumR} \\
		\frac{1}{k} \sum_{i=1}^k {D(k \tilde{R}_i)} \leq \bar{d}, \label{eqn:avgDhatR<d}
	\end{IEEEeqnarray}
	where $R_i$ was defined in \eqref{eqn:def R}.
	
	\item\label{itm:decR}
	In this step, we find a relation between $\tilde{R}_i'$ and $G^{\rm eff}_k$ (which is defined later).
	The sequence $\tilde{R}'_i$ is generated from $\tilde{R}_i$ by increasing $\tilde{R}_1$ such that the sequence has the same total rate as the total effective rate of the sequence $R_i$ and by causally shifting the rates $\tilde{R}_i$.
	Also, the sequence $\tilde{R}'_i$ satisfies the distortion constraint. \\
	Formally, for any $k\in\mathbb{N}$, there exists a sequence $\tilde{R}'_i$, for $i=1,\ldots,k$, such that
	\begin{IEEEeqnarray}{l}
		\tilde{R}'_1\geq\cdots\geq \tilde{R}'_k, \label{eqn:R1>Rn} \\
		\sum_{i=1}^j{\tilde{R}'_i}
		\geq G^{\rm eff}_k\left(\frac{j}{k}\right),
		\qquad j=1,\ldots,k-1, \label{eqn:sumRt''>GL} \\
		\sum_{i=1}^k{\tilde{R}'_i}
		= G^{\rm eff}_k(1), \label{eqn:totRt''=GL1-GL0} \\
		\frac{1}{k} \sum_{i=1}^k {D(k \tilde{R}'_i)}
		\leq \bar{d}, \label{eqn:avgDhatR''<d}
	\end{IEEEeqnarray}
	where
	\begin{equation} \label{eqn:GLk}
		G^{\rm eff}_k(\alpha)
		:= \max\left\{0, G(\alpha) - \max_{j\in\{0,\ldots,k\}}\left(G\left(\frac{j}{k}\right) - L\left(\frac{j}{k}\right)\right)\right\}.
	\end{equation}

	\item\label{itm:existConcG}
	In this step, we show that $\tilde{R}_i'$ majorizes the samples of the concave hull of $G^{\rm eff}_k(i/k)$ (denoted by $\hat{G}_k(i/k)$ which is defined later).
	Then, using the majorization inequality (see Lemma \ref{lmm:MajIneq}), we show that the concave hull satisfies the distortion constraint. \\
	Formally, we define a $\crdf$ function $G_k$ as an approximation of $G^{\rm eff}_k$ as following:
	\begin{equation} \label{eqn:Gk}
		G_k(\alpha)
		:= G^{\rm eff}_k\left( \frac{\floor{\alpha k}}{k} \right),
		\qquad \alpha\in[0,1].
	\end{equation}
	Further, we denote the envelope of $G_k$ by $\hat{G}_k$ w.r.t. Definition \ref{def:ConvHull}.
	Also, we define
	\begin{equation} \label{eqn:bRi}
		\hat{R}_i
		:= \hat{G}_k \left( \frac{i}{k} \right) 
		- \hat{G}_k \left( \frac{i-1}{k} \right),
		\qquad i = 1, \ldots, k.
	\end{equation}
	Then, we have
	\begin{equation} \label{eqn:avgD(bR)<d}
		\frac{1}{k} \sum_{i=1}^k {D(k \hat{R}_i)}
		\leq \frac{1}{k} \sum_{i=1}^k {D(k \tilde{R}'_i)}
		\leq \bar{d}.
	\end{equation}

	\item\label{itm:ConvDiscCon}
	In this step, we use the regularity conditions of $\crdf$ and $\cldf$ to prove that the result is correct when $k\to\infty$.
	With $\hat{R}_i$ defined in \eqref{eqn:bRi}, the following limit is valid:
	\begin{equation} \label{eqn:bGk->bG}
		\lim_{k\to\infty}{\frac{1}{k} \sum_{i=1}^k {D(k \hat{R}_i)}}
		=\int_0^1 {D\left(\frac{\ud \hat{G}^{\rm eff}}{\ud \alpha}(\alpha) \right) \ud \alpha}.
	\end{equation}
\end{enumerate}

\emph{\bf Proof of Step \ref{itm:ClasConv})}:
\emph{Proof of \eqref{eqn:G-Gj/k>sumRhat}}:
For any $k,n\in\mathbb{N}$ and $j\in\{1, \ldots, k-1\}$, we have
\begin{IEEEeqnarray}{rCl}
	n k \sum_{i=j+1}^k{R_i}
	&\stackrel{(a)}{\geq}& \sum_{i=j+1}^k{\H{M_i}}
	\geq \Hb{M_{j+1}^k}
	\geq \Hb{M_{j+1}^k \vert M^j}
	\geq \Ib{X^{nk}}{M_{j+1}^k \vert M^j} \label{eqn:sumR>I+I1}\\
	&=& \Ib{X_{n j+1}^{nk}}{M_{j+1}^k \vert M^j}
	+ \Ib{X^{n j}}{M_{j+1}^k \vert X_{n j+1}^{n k}, M^j}, \label{eqn:sumR>I+I}
\end{IEEEeqnarray}
where $(a)$ follows from the definition of $R_i$ (see Definition \ref{def:Code}).
We define, for $i\in\{1,\ldots,k\}$,
\begin{equation} \label{eqn:Rtn}
	\tilde{R}_i^{(n)}
	:= \frac{1}{k} R\big(\Eb{d(X_{n (i-1)+1}^{n i}, \hat{X}_{n (i-1)+1}^{n i})}\big),
	\qquad \tilde{R}_i := \liminf_{n\to\infty}{\tilde{R}_i^{(n)}},
\end{equation}
which $R(\cdot)$ is the rate-distortion function for $X \sim p(x)$ and distortion function $d$ (see Definition \ref{def:classical R(D)}).
Later, we show that for $j=1,\ldots,k-1$,
\begin{IEEEeqnarray}{l}
	\Ib{X_{n j + 1}^{n k}}{M_{j+1}^k \vert M^j}
	\geq n k \sum_{i=j+1}^k{\tilde{R}_i^{(n)}}, \label{eqn:I>sumRt}
	\;\footnote{\text{This inequality is also valid for $j=0$ using the notation $M^0=\emptyset$.}}\\
	\Ib{X^{n j}}{M_{j+1}^k \vert M^j, X_{n j + 1}^{n k}}
	\geq n k \max\left\{0, \sum_{i=1}^j{\tilde{R}_i^{(n)}} - L\left(\tfrac{j}{k}\right)\right\}. \label{eqn:I<max0,R-L}
\end{IEEEeqnarray}
Hence, from \eqref{eqn:sumR>I+I}, \eqref{eqn:I>sumRt}, and \eqref{eqn:I<max0,R-L}, we obtain
\begin{equation*} \label{eqn:sumR>sumRtn+max0}
	\sum_{i=j+1}^k{R_i}
	\geq \sum_{i=j+1}^k{\tilde{R}_i^{(n)}}
	+ \max\left\{0, \sum_{i=1}^j{\tilde{R}_i^{(n)}} - L\left(\tfrac{j}{k}\right)\right\} 
	= \sum_{i=1}^k{\tilde{R}_i^{(n)}}
	- \min\left\{\sum_{i=1}^j{\tilde{R}_i^{(n)}}, L\left(\tfrac{j}{k}\right)\right\}.
\end{equation*}
Thus, \eqref{eqn:G-Gj/k>sumRhat} is obtained from \eqref{eqn:Rtn}, by taking $\liminf$ of both sides of the inequality as $n\to\infty$.

\emph{Proof of \eqref{eqn:sumRt<L1,sumR}}:
We obtain that
\begin{equation} \label{eqn:sumRt<sumRnew}
	n k \sum_{i=1}^k{R_i}
	\stackrel{(a)}{\geq} \Ib{X^{nk}}{M^k}
	\stackrel{(b)}{\geq} nk \sum_{i=1}^k{\tilde{R}_i^{(n)}},
\end{equation}
where $(a)$ follows from the similar steps of \eqref{eqn:sumR>I+I1} for $j=0$ by using the notation $M^0 = \emptyset$ and $(b)$ follows from \eqref{eqn:I>sumRt}.
Further, we can write
\begin{equation} \label{eqn:sumRt<L1}
	nk\sum_{i=1}^k \tilde{R}_i^{(n)}
	\stackrel{(a)}{\leq} \Ib{X^{nk}}{M^k}
	\stackrel{(b)}{\leq} nk L(1).
\end{equation}
where $(a)$ follows from \eqref{eqn:I>sumRt} and $(b)$ follows from \eqref{eqn:mainI<L}.
Hence, from \eqref{eqn:sumRt<sumRnew} and \eqref{eqn:sumRt<L1}, we obtain
\begin{equation*} \label{eqn:sumRtn<L1,sumR}
	\sum_{i=1}^k \tilde{R}_i^{(n)}
	\leq \min\left\{L(1), \sum_{i=1}^k{R_i}\right\},
\end{equation*}
Thus, \eqref{eqn:sumRt<L1,sumR} is obtained from \eqref{eqn:Rtn}, by taking $\liminf$ of both sides of the inequality as $n\to\infty$.

\emph{Proof of \eqref{eqn:avgDhatR<d}}:
From Definition \ref{def:classical R(D)}, we have
\begin{equation} \label{eqn:avgDn<d}
	\frac{1}{k} \sum_{i=1}^k {D(k \tilde{R}_i^{(n)})}
	\stackrel{(a)}{\leq} \frac{1}{k} \sum_{i=1}^k{\Eb{d(X_{n (i-1)+1}^{n i}, \hat{X}_{n (i-1)+1}^{n i})}}
	\stackrel{(b)}{=} \E{d(X^{n k},\hat{X}^{n k})},
\end{equation}
where $D(\cdot)$ is the distortion-rate function;
$\tilde{R}_i^{(n)}$ was defined in \eqref{eqn:Rtn};
$(a)$ follows from the definition of $D(\cdot)$;
and $(b)$ follows from \eqref{eqn:partiald}.
From Corollary \ref{cor:R(D)convex}, $D(\cdot)$ is continuous and non-increasing.
As a result, we obtain
$
	\limsup_{n\to\infty}{D(k \tilde{R}_i^{(n)})}
	= D(k \liminf_{n\to\infty}{\tilde{R}_i^{(n)}})
	= D(k \tilde{R}_i).
$
Hence, by taking $\limsup$ from both sides of \eqref{eqn:avgDn<d}, we obtain that
\begin{equation*}
	\frac{1}{k} \sum_{i=1}^k{D(k \tilde{R}_i)}
	\leq \limsup_{n\to\infty}\E{d(X^{n k},\hat{X}^{n k})}
	\stackrel{(a)}{\leq} \bar{d},
\end{equation*}
where $(a)$ follows from the fact that, in the converse part, we assume that $G$ is achievable (see Definition \ref{def:Achievable R(D)}).

Thus, Step \ref{itm:ClasConv} is proved.
Now, it only remains to prove \eqref{eqn:I>sumRt} and \eqref{eqn:I<max0,R-L}.

\emph{Proof of \eqref{eqn:I>sumRt}}:
The claim follows from the following sequence of inequalities:
\begin{IEEEeqnarray}{rCl}
	\Ib{X_{n j+1}^{n k}}{M_{j+1}^k \,\big\vert\, M^j}
	&=& \Ib{X_{n j + 1}^{n k}}{M^k} \label{eqn:XindepM}\\
	&=& \Ib{X_{n j + 1}^{n k}}{M^k, \hat{X}_{n j + 1}^{n k}}
	\geq \Ib{X_{n j + 1}^{n k}}{\hat{X}_{n j + 1}^{n k}} \label{eqn:Xhat=f(M)}\\
	&=& \sum_{i=nj+1}^{n k}{\H{X_i} - \Hb{X_i \big\vert X_{n j+1}^{i-1}, \hat{X}_{nj+1}^{n k}}}
	\geq \sum_{i=nj+1}^{n k}{\Ib{X_i}{\hat{X}_i}} \label{eqn:Xkiid}\\
	&\geq& \sum_{i=nj+1}^{n k}{R\big(\Eb{d(X_i, \hat{X}_i)}\big)}
	= \sum_{i=j+1}^k
	{\sum_{\ell=n (i-1)+1}^{n i}{R\big(\Eb{d(X_\ell, \hat{X}_\ell)}\big)}} \label{eqn:I>R(Ed)}\\
	&\geq& \sum_{i=j+1}^k
	{n R\big(\Eb{d(X_{n (i-1)+1}^{n i}, \hat{X}_{n (i-1)+1}^{n i})}\big)}
	= \sum_{i=j+1}^k
	{n k \tilde{R}_i^{(n)}}, \label{eqn:G-Ga>sumkRk}
\end{IEEEeqnarray}
where \eqref{eqn:XindepM} follows since $X^{nj}$ is independent of $X_{nj+1}^{n k}$, as a result, according to Definition \ref{def:Code}, $M^j$ is independent of $X_{nj+1}^{n k}$;
the equality of \eqref{eqn:Xhat=f(M)} follows because $\hat{X}_{nj+1}^{n k}$ is a function of $M^k$;
the equality of \eqref{eqn:Xkiid} follows since the sequence $\{X_i\}$ is $\IID$;
the inequality of \eqref{eqn:I>R(Ed)} follows based on the definition of $R(\cdot)$ \cite[p. 307]{Cover06};
and the inequality of \eqref{eqn:G-Ga>sumkRk} follows from the convexity of $R(D)$ (see Corollary \ref{cor:R(D)convex}).

\emph{Proof of \eqref{eqn:I<max0,R-L}}:
The claim follows from the following sequence of inequalities:
\begin{IEEEeqnarray}{rCl}
	\Ib{X^{n j}}{M_{j+1}^k \vert M^j, X_{n j + 1}^{n k}}
	&=& \Hb{X^{nj} \vert M^j, X_{n j + 1}^{n k}} 
	- \Hb{X^{nj} \vert M^k, X_{n j + 1}^{n k}} \nonumber\\
	&=& \Hb{X^{nj} \vert M^j} 
	- \Hb{X^{nj} \vert M^k, X_{n j + 1}^{n k}}
	\geq \Hb{X^{nj} \vert M^j} 
	- \Hb{X^{nj} \vert M^k} \label{M=f(X1),X1indX2} \IEEEeqnarraynumspace\\
	&=& \Hb{X^{nj} \vert M^j} 
	- \Hb{X^{nj} \vert M^k, \hat{X}^{nj}} \label{eqn:Xh=f(M)} \\
	&\geq& \Hb{X^{nj} \vert M^j}
	 - \Hb{X^{nj} \vert \hat{X}^{nj}}
	= -\Ib{X^{nj}}{M^j} 
	+ \Ib{X^{nj}}{\hat{X}^{nj}} \nonumber\\
	&\geq& - n k L\left(\tfrac{j}{k}\right) 
	+ \Ib{X^{nj}}{\hat{X}^{nj}} \label{eqn:-I>-L} \\
	&\geq& - n k L\left(\tfrac{j}{k}\right) 
	+ n k \sum_{i=1}^j{\tilde{R}_i^{(n)}}, \label{eqn:I<R(E[d])}
\end{IEEEeqnarray}
where, the equality of \eqref{M=f(X1),X1indX2} follows since $X^{nj}$ is independent of $X_{nj+1}^{n k}$, as a result, according to Definition \ref{def:Code}, $M^j$ is independent of $X_{nj+1}^{n k}$;
\eqref{eqn:Xh=f(M)} follows from the fact that $\hat{X}^{nj}$ is a function of $M^k$ (see Definition \ref{def:Code});
\eqref{eqn:-I>-L} follows from \eqref{eqn:mainI<L};
\eqref{eqn:I<R(E[d])} is obtained following the similar steps as inequalities \eqref{eqn:Xkiid}, \eqref{eqn:I>R(Ed)}, and \eqref{eqn:G-Ga>sumkRk}.

\emph{\bf Proof of Step \ref{itm:decR})}:
We define $\{S_i\}_{i=1}^k$ as the sorted permutation of $\{\tilde{R}_i\}_{i=1}^k$ in descending order.
Hence, we obtain
\begin{IEEEeqnarray}{l}
	S_1 \geq \cdots \geq S_k, \label{eqn:yr1>yrn} \\
	\sum_{i=1}^j {S_i} 
	\geq \sum_{i=1}^j {\tilde{R}_i}
	\qquad j=1,\ldots,k-1, \label{eqn:sumyi<sumRt} \\
	\sum_{i=1}^k {S_i} 
	= \sum_{i=1}^k {\tilde{R}_i}, \label{eqn:sumSi=sumRt} \\
	\sum_{i=1}^k {D(k S_i)} 
	= \sum_{i=1}^k {D(k \tilde{R}_i)}. \label{eqn:avgDhatyr<d}
\end{IEEEeqnarray}
Next, we define the sequence $\tilde{R}'_j$, for $j=1,\ldots,k$, as
\begin{equation} \label{eqn:def:tildeR'i}
	\tilde{R}'_j =
	\begin{cases}
		S_1 + G^{\rm eff}_k(1) - \sum_{i=1}^k S_i & j = 1, \\
		S_j & j=2, \ldots, k.
	\end{cases}
\end{equation}
Later, we show that
\begin{equation} \label{eqn:ctilde>cnormal}
	\sum_{i=1}^k{\tilde{R}_i}
	\leq G(1) - \max_{j\in\{0,\ldots,k\}}G\left(\tfrac{j}{k}\right) 
	- L\left(\tfrac{j}{k}\right)
	= G^{\rm eff}_k(1).
\end{equation}
Hence, $\tilde{R}'_1 \geq S_1$ due to \eqref{eqn:sumSi=sumRt};
as a result, $\{\tilde{R}'_j\}_{j=1}^k$ is a valid sequence of rates.

\emph{Proof of \eqref{eqn:R1>Rn}}:
From \eqref{eqn:yr1>yrn} and \eqref{eqn:def:tildeR'i}, we obtain that $S_1\geq\tilde{R}'_2\geq\ldots\geq\tilde{R}'_k$.
Thus, \eqref{eqn:R1>Rn} follows from the fact that $\tilde{R}'_1 \geq S_1$.

\emph{Proof of \eqref{eqn:sumRt''>GL}}:
We can write, for $j=1,\ldots,k-1$,
\begin{equation} \label{eqn:sumRt'>GL1-sumRt_}
	\sum_{i=1}^j \tilde{R}'_i
	\stackrel{(a)}{=} G_k^{\rm eff}(1) - \sum_{i=1}^k S_i
	+ \sum_{i=1}^j S_i
	= G_k^{\rm eff}(1) - \sum_{i=j+1}^k S_i
	\stackrel{(b)}{\geq} G_k^{\rm eff}(1) - \sum_{i=j+1}^{k} \tilde{R}_i,
\end{equation}
where $(a)$ follows from \eqref{eqn:def:tildeR'i}
and $(b)$ follows from \eqref{eqn:sumyi<sumRt} and \eqref{eqn:sumSi=sumRt}.
From \eqref{eqn:GLk}, we obtain that
\begin{equation} \label{eqn:GL1-GLj>sumRt_}
	G_k^{\rm eff}(1) - G_k^{\rm eff}\left(\tfrac{j}{k}\right)
	\stackrel{(a)}{=} G(1) - c - \max\left\{0, G\left(\tfrac{j}{k}\right) - c \right\}
	= \min\left\{G_k^{\rm eff}(1) , G(1) - G\left(\tfrac{j}{k}\right) \right\}
	\stackrel{(b)}{\geq} \sum_{i=j+1}^{k} \tilde{R}_i,
\end{equation}
where $(a)$ follows from \eqref{eqn:ctilde>cnormal} with $c := \max_{j\in\{0,\ldots,k\}}G\left({j}/{k}\right) - L\left({j}/{k}\right)$;
$(b)$ follows from \eqref{eqn:ctilde>cnormal} because $\sum_{i=1}^k \tilde{R}'_i \geq \sum_{i=j+1}^k \tilde{R}'_i$ and from \eqref{eqn:G-Gj/k>sumRhat} because $G(1) - G\left({j}/{k}\right)\geq \sum_{i=j+1}^k{\tilde{R}_i}$.
Thus, \eqref{eqn:sumRt''>GL} follows from \eqref{eqn:sumRt'>GL1-sumRt_} and \eqref{eqn:GL1-GLj>sumRt_}.

\emph{Proof of \eqref{eqn:totRt''=GL1-GL0}}:
It follows from \eqref{eqn:def:tildeR'i}.

\emph{Proof of \eqref{eqn:avgDhatR''<d}}:
Since the distortion-rate function $D(\cdot)$ is non-increasing (see Remark \ref{cor:R(D)convex}), $D(k\tilde{R}'_1) \leq D(k S_1)$ because $\tilde{R}'_1 \geq S_1$.
Besides, for $i=2,\ldots,k$, $D(k \tilde{R}'_i) = D(k S_i)$; as a result, \eqref{eqn:avgDhatR''<d} follows from \eqref{eqn:avgDhatyr<d}.

Thus, Step \ref{itm:decR} is proved.
Now, it only remains to prove \eqref{eqn:ctilde>cnormal}.

\emph{Proof of \eqref{eqn:ctilde>cnormal}}:
To prove the equality, we have that $G(1) \geq G(j/k) \geq G(j/k) - L(j/k)$ because $G(\cdot)$ is non-decreasing and $L(\alpha) \geq 0$ for all $\alpha\in [0,1]$.
Hence, $G(1) \geq \max_{j\in\{0,\ldots,k\}} G(j/k) - L(j/k)$; as a resut, the equality follows from \eqref{eqn:G-Gj/k>sumRhat}.

To prove the inequality, we need to show that
$
	G(1) - \sum_{i=1}^k{\tilde{R}_i}
	\geq G\left({j}/{k}\right) 
	- L\left({j}/{k}\right),
$
for all $j=0,\ldots,k$.
For $j=k$, utilizing \eqref{eqn:sumRt<L1,sumR}, we obtain
\begin{equation} \label{eqn:G1-sumRt j=k}
	\sum_{i=1}^k{\tilde{R}_i}
	\leq L(1)
	\Longrightarrow G(1) - \sum_{i=1}^k{\tilde{R}_i}
	\geq G(1) - L(1).
\end{equation}
For $j=1,\ldots,k-1$, from \eqref{eqn:G-Gj/k>sumRhat} and Definition \ref{def:Code}, we obtain that,
\begin{equation} \label{eqn:G1-sumRt j!=k}
	G(1) - G\left(\tfrac{j}{k}\right)
	\geq \sum_{i=1}^k{\tilde{R}_i} - L\left(\tfrac{j}{k}\right)
	\Longrightarrow
	G(1) - \sum_{i=1}^k{\tilde{R}_i}
	\geq G\left(\tfrac{j}{k}\right) 
	- L\left(\tfrac{j}{k}\right).
\end{equation}
For $j=0$, first, note that
\begin{equation} \label{eqn:G1>sumRt_}
	G(1) = \sum_{i=1}^k{R_i}
	\stackrel{(a)}{\geq} \sum_{i=1}^k{\tilde{R}_i}
	\Longrightarrow G(1) - \sum_{i=1}^k{\tilde{R}_i} \geq 0,
\end{equation}
where $(a)$ follows from \eqref{eqn:sumRt<L1,sumR}.
Hence, for $j=0$, it follows from \eqref{eqn:G1>sumRt_} and the fact that $G(0) - L(0)=0$, which follows from the zero initial value property of $G$ and $L$ (see Definitions \ref{def:CRDF} and \ref{def:CLDF}).

\emph{\bf Proof of Step \ref{itm:existConcG})}:
Using $\tilde{R}'_i$, $i=1,\ldots,k$, we define a continuous and piece-wise linear function $\tilde{G}_k$ with slope of $k \tilde{R}'_i$ for $\alpha\in [(i-1)/k, i/k)$ and $\tilde{G}_k(1) = G^{\rm eff}_k(1)$.
Formally,
\begin{equation} \label{eqn:tildeGk}
	\tilde{G}_k(\alpha)
	:=G^{\rm eff}_k(1) - \sum_{i=\ceil{\alpha k}+1}^k{\tilde{R}'_i}
	+ \tilde{R}'_{\ceil{\alpha k}} \left( \alpha k - \ceil{\alpha k} \right),
	\qquad \alpha\in[0,1],
\end{equation}
with the abuse of notation that if $\ceil{\alpha k} + 1 > k$, the summation is assumed to be $0$ and $\tilde{R}'_0$ is an arbitrary finite number.
Hence, from \eqref{eqn:sumRt''>GL} and \eqref{eqn:totRt''=GL1-GL0}, we obtain that
\begin{equation} \textstyle \label{eqn:tildeGk>G}
	\tilde{G}_k(0) = 0,
	\quad\tilde{G}_k(1) = G^{\rm eff}_k(1),
	\quad\tilde{G}_k\left(\frac{j}{k}\right)
	\geq G^{\rm eff}_k\left(\frac{j}{k}\right),
	\;\forall j = 1,\ldots,k-1.
\end{equation}
Hence, from \eqref{eqn:tildeGk>G}, the definitions of $G_k$ and $\tilde{G}_k$ in \eqref{eqn:Gk} and \eqref{eqn:tildeGk}, respectively, and the fact that $\tilde{G}_k^{\rm eff}$ is non-decreasing, we have that
\begin{equation} \label{eqn:tGka>=Gka}
	\tilde{G}_k(0) = G_k(0),
	\quad\tilde{G}_k(1) = G_k(1),
	\quad\tilde{G}_k(\alpha) \geq G_k(\alpha),
	\;\forall\alpha \in (0,1).
\end{equation}
Since $\tilde{R}'_1 \geq \cdots \geq \tilde{R}'_n\geq 0$, we obtain that the derivative of $\tilde{G}_k$ is non-increasing and non-negative; as a result, $\tilde{G}_k$ is concave and non-decreasing.
Hence, from the definition of the envelope (see Definition \ref{def:ConvHull}), and the fact that $\tilde{G}_k$ is concave, we obtain
\begin{equation} \label{eqn:tildeGk>convG}
	\tilde{G}_k(\alpha)
	\geq \hat{G}_k(\alpha),
	\qquad \forall \alpha\in [0,1],
\end{equation}
where $\hat{G}_k$ is the envelope of $G_k$.
By recalling Lemma \ref{lmm:maxf=f(b)}, we obtain that
\begin{equation} \label{eqn:convG01=tG01}
	\hat{G}_k(0) = G_k(0) = 0,
	\qquad\hat{G}_k(1) = G_k(1) = G_k^{\rm eff}(1).
\end{equation}
Hence, utilizing \eqref{eqn:tGka>=Gka}, \eqref{eqn:tildeGk>convG}, and \eqref{eqn:convG01=tG01}, we have
\begin{equation} \label{eqn:tGka>=convGka}
	\tilde{G}_k(0) = \hat{G}_k(0) = 0,
	\quad \tilde{G}_k(1) = \hat{G}_k(1) = G_k^{\rm eff}(1),
	\quad\tilde{G}_k(\alpha) \geq \hat{G}_k(\alpha),
	\;\forall\alpha \in (0,1).
\end{equation}
Therefore, for all $j=1,\ldots,k-1$, we obtain
\begin{equation}
	\sum_{i=1}^j {\hat{R}_i}
	\stackrel{(a)}{=} \hat{G}_k\left(\tfrac{j}{k}\right)
	- \hat{G}_k(0)
	\stackrel{(b)}{=} \hat{G}_k\left(\tfrac{j}{k}\right)
	\stackrel{(c)}{\leq} \tilde{G}_k\left(\tfrac{j}{k}\right)
	\stackrel{(d)}{=} \sum_{i=1}^j {\tilde{R}'_i}, \label{eqn:Rb Maj Rtild}
\end{equation}
where $(a)$ follows from the definition of $\hat{R}_i$ in \eqref{eqn:bRi};
$(b)$ and $(c)$ follow from \eqref{eqn:tGka>=convGka}; 
and $(d)$ follows from the definition of $\tilde{G}_k$ in \eqref{eqn:tildeGk} and the fact that $\tilde{G}_k(0) = 0$ in \eqref{eqn:tildeGk>G}.
Simillarly, we obtain
\begin{equation} \label{eqn:conG1-conG0>StR}
	\sum_{i=1}^k {\hat{R}_i}
	= \hat{G}_k(1)
	- \hat{G}_k(0)
	= \hat{G}_k(1)
	\stackrel{(a)}{=} \sum_{i=1}^k {\tilde{R}'_i},
\end{equation}
where $(a)$ follows from \eqref{eqn:tGka>=convGka} and the definition of $\tilde{G}_k$ in \eqref{eqn:tildeGk}.
Since $\hat{G}_k$ is concave, we have 
\begin{equation} \label{eqn:bRdec}
	\hat{R}_1\geq\cdots\geq \hat{R}_k.
\end{equation}
Therefore, utilizing the convexity of $D(R)$ (see Remark \ref{cor:R(D)convex}) and the majorization inequality (see Lemma \ref{lmm:MajIneq}), we further obtain from \eqref{eqn:R1>Rn}, \eqref{eqn:Rb Maj Rtild}, \eqref{eqn:conG1-conG0>StR}, and \eqref{eqn:bRdec} that
$
	\sum_{i=1}^k {D(k \hat{R}_i)}
	\leq \sum_{i=1}^k {D(k \tilde{R}'_i)}.
$
Hence, \eqref{eqn:avgD(bR)<d} follows from \eqref{eqn:avgDhatR<d} and the derivation of Step \ref{itm:existConcG} is complete.

\emph{\bf Proof of Step \ref{itm:ConvDiscCon})}:
The function $\hat{G}_k$ is continuous and piece-wise linear because it is the concave hull of the points $G^{\rm eff}_k(i/k)$, for $i=0,\ldots,k$.
Hence, the possible indifferentiable points are located at $i/k$ for $i=1,\ldots,k-1$ and the derivative is $k\hat{R}_i$ for $\alpha\in ((i-1)/k, i / k)$.
Formaly,
\begin{equation} \label{eqn:tildeGk}
	\hat{G}_k(\alpha)
	:=G^{\rm eff}_k(1) - \sum_{i=\ceil{\alpha k}+1}^k{\hat{R}_i}
	+ \hat{R}_{\ceil{\alpha k}} \left( \alpha k - \ceil{\alpha k} \right),
	\qquad \alpha\in[0,1],
\end{equation}
with the abuse of notation that if $\ceil{\alpha k} + 1 > k$, the summation is assumed to be $0$ and $\hat{R}_0$ is an arbitrary finite number.
Hence,
\begin{equation}
	\frac{1}{k} \sum_{i=1}^k {D(k \hat{R}_i)} 
	= \int_0^1{D\left(\tfrac{\ud \hat{G}_k}{\ud \alpha}(\alpha) \right) \ud\alpha}.\label{equality_step4}
\end{equation}
If we show that the derivative of $\hat{G}_k$ tends (point-wise) to the derivative of $\hat{G}^{\rm eff}$ almost everywhere in $[0,1]$, then, using the dominated convergence theorem \cite[Lemma 5.10]{Gray09} and the fact that $D(\cdot)$ is bounded and continuous (see Remark \ref{cor:R(D)convex}), \eqref{eqn:bGk->bG} is proved.
To do so, we define, for $\rho\geq 0$,
\begin{equation} \label{eqn:B(R)}
	B(\rho) := \sup_{\beta\in[0,1]}{G^{\rm eff}(\beta) - \rho \beta},
	\quad B^{\rm eff}_k := \sup_{\beta\in[0,1]}{G^{\rm eff}_k(\beta) - \rho \beta},
	\quad B_k(\rho) := \sup_{\beta\in[0,1]}{G_k(\beta) - \rho \beta},
\end{equation}
where $G_k(\cdot)$ was defined in \eqref{eqn:Gk}.
Then, from Lemma \ref{lmm:Dconvf}, and the fact that $G^{\rm eff}(1) \leq G_k(1) = G^{\rm eff}_k(1) \leq G(1)$, we have that for a given $\alpha \in (0,1) \setminus (\mathcal{E} \cup \mathbb{Q})$, we have
\begin{IEEEeqnarray}{l} 
	\rho^* := \frac{\ud\hat{G}^{\rm eff}}{\ud \alpha}(\alpha) 
	= \argmin_{\rho\in [0 , G(1)/\alpha]}{\rho \alpha + B(\rho)}, \label{eqn:rho*}\\
	\rho^*_k := \frac{\ud\hat{G}_k}{\ud \alpha}(\alpha) 
	= \argmin_{\rho\in [0 , G(1)/\alpha]}{\rho \alpha + B_k(\rho)}, \label{eqn:rhok}
\end{IEEEeqnarray}
where $\mathcal{E}$ is a set of the point at which the derivative of $\hat{G}^{\rm eff}$ does not exist which is countable from Lemma \ref{lmm:Dconvf}.
The reason that we removed the set of rational numbers is that it can be proved that the derivatives of $G_k$ are uniquely defined at all the irrational points and for all $k\in\mathbb{N}$.
Hence, $\mathcal{E} \cup \mathbb{Q}$ is countable and the dominated convergence theorem \cite[Lemma 5.10]{Gray09} applies for this case. \\
Next, we prove that the limit of $\rho^*_k$ exists and it converges to $\rho^*$ as $k\to\infty$.
First, we show the existense of the limit.
To this end, recall \eqref{eqn:rhok} that, for all $k\in\mathbb{N}$, $\rho^*_k$ is in $[0 , G(1)/\alpha]$; 
as a result, it is bounded and $\limsup_{k\to\infty}{\rho^*_k}$ and $\liminf_{k\to\infty}{\rho^*_k}$ exist.
So, there exist subsequences of $\{\rho^*_k\}_{k=1}^\infty$ converging to $\limsup_{k\to\infty}{\rho^*_k}$ and $\liminf_{k\to\infty}{\rho^*_k}$, respectively. 
Hence, it suffices to show that, for any convergent subsequence, the limit of the subsequence is $\rho^*$ because in this case $\limsup_{k\to\infty}{\rho^*_k} = \liminf_{k\to\infty}{\rho^*_k} = \rho^*$ and the statement is proved.\\
Thus, without loss of generality, we assume that the convergent sequence is $\rho^*_k$ itself, which it converges to $\hat\rho$:
\begin{equation} \label{rhohat}
	\hat\rho = \lim_{k\to\infty}{\rho_k^*}.
\end{equation}
Therefore, utilizing \eqref{eqn:rho*}, it suffices to show that
\begin{equation} \label{eqn:hatRho=Rho*}
	\hat\rho \alpha + B(\hat\rho) \leq \rho\alpha + B(\rho),
	\qquad \forall \rho\in[0,G(1)/\alpha];
\end{equation}
as a result, since $\rho^*$ is unique due to the existence and uniqueness of the derivative of $\hat{G}(\alpha)$, $\hat\rho = \rho^*$.
From \eqref{eqn:rhok}, we have that for all $\rho \geq 0$,
\begin{equation} \label{eqn:rho + B_k<rho + B_L}
	\rho^*_k \alpha + B_k(\rho_k^*)
	\leq \rho \alpha + B_k(\rho)
	\stackrel{(a)}{\leq} \rho \alpha + B^{\rm eff}_k(\rho)
	\leq \rho \alpha + B(\rho) + \epsilon_k,
\end{equation}
where the $(a)$ follows from \eqref{eqn:B(R)}, and the fact that $G_k(\alpha)\leq G^{\rm eff}_k(\alpha)$ because $G(\cdot)$ and, as a result, $G^{\rm eff}_k(\cdot)$, are non-decreasing.
Later, we show that
\begin{equation} \label{eqn:epsilon -> 0}
	\lim_{k\to\infty} \epsilon_k = 0.
\end{equation}
Next, by taking the limit from both sides of the inequality and utilizing \eqref{rhohat}, we obtain that 
\begin{equation*}
	\hat \rho \alpha + \lim_{k\to\infty}{B_k(\rho_k^*)}
	\leq \rho \alpha + B(\rho).
\end{equation*}
Therefore, to prove \eqref{eqn:hatRho=Rho*}, it suffices to show that 
\begin{equation} \label{eqn:BkrhoK->Bhrho}
	\lim_{k\to\infty}{B_k(\rho_k^*)}
	= B(\hat\rho).
\end{equation}
To this end, we make use of the following result which we prove later:
\begin{equation} \label{eqn:B-Bk<r/k}
	0 \leq B(\rho) - B_k(\rho) \leq \frac{\rho}{k},
	\qquad \forall \rho\geq 0, k\in\mathbb{N}.
\end{equation}
Thus, we have that
\begin{equation*}
	\left\lvert B_k(\rho_k^*) - B(\hat\rho) \right\rvert
	\leq \left\lvert B_k(\rho_k^*) - B(\rho_k^*) \right\rvert
	+ \left\lvert B(\rho_k^*) - B(\hat\rho) \right\rvert
	\stackrel{(a)}{\leq} \frac{\rho_k^*}{k} 
	+ \left\lvert B(\rho_k^*) - B(\hat\rho) \right\rvert,
\end{equation*}
where $(a)$ follows from \eqref{eqn:B-Bk<r/k}.
Next, since $B(\rho)$ is convex and bounded, over $\rho\in [0,G(1)/\alpha]$, it is also continuous in $[0,G(1)/\alpha]$ (for details, see \eqref{eqn:f0<ba<f1} in Lemma \ref{lmm:Dconvf}); as a result, $\left\lvert B(\rho_k^*) - B(\hat\rho) \right\rvert$ can become small enough for large $k$.
Therefore, because $\rho^*_k \in [0,G(1)/\alpha]$ (see \eqref{eqn:rhok}), \eqref{eqn:BkrhoK->Bhrho} is proved.
Therefore, it only remains to prove \eqref{eqn:epsilon -> 0} and \eqref{eqn:B-Bk<r/k}.

\emph{Proof of \eqref{eqn:epsilon -> 0}}:
From \eqref{eqn:Gk}, it is obtained that $G_k(\alpha) \leq G^{\rm eff}_k(\alpha)$ for $\alpha\in[0,1]$.
Utilizing the definitions of $G^{\rm eff}$ and $G^{\rm eff}_k$, in \eqref{eqn:GL} and \eqref{eqn:GLk}, respectively, we obtain that
\begin{equation*}
	0 \leq G^{\rm eff}_k(\alpha) - G^{\rm eff}(\alpha)
	\leq \epsilon_k,
\end{equation*}
where
\begin{equation*}
	\epsilon_k := \sup_{\beta\in[0,1]} F(\beta)
	- \max_{j\in\{0,\ldots,k\}} F\left(\tfrac{j}{k}\right),
	\qquad F(\beta) := G(\beta) - L(\beta).
\end{equation*}
To prove \eqref{eqn:epsilon -> 0}, it is sufficient to show that
\begin{equation} \label{eqn:epsilonk->0 Analytics}
 \forall\delta > 0, ~\exists K\in\mathbb{N}\colon k \geq K \Rightarrow\epsilon_k \leq \delta.
\end{equation}
If $\sup_{\beta\in[0,1]}F(\beta) = F(1)$, then it is clear that $\epsilon_k = 0$ for all $k$; as a result claim is proved.
Now, we study the case that $\sup_{\beta\in[0,1]}F(\beta) > F(1)$.
Function $F$ is bounded because $G$ and $L$ are bounded.
Therefore,
\begin{equation} \label{eqn:F(b)-F(b')<d/2}
	\forall\delta > 0, \exists\beta'\in[0,1) \colon \sup_{\beta\in[0,1]}F(\beta) - F(\beta') 
	\leq \frac{\delta}{2}.
\end{equation}
Function $F$ is right continuous because both $G$ and $L$ are right continuous (see Definitions \ref{def:CRDF} and \ref{def:CLDF}).
Hence,
\begin{equation*}
	\forall\delta>0,~\exists\gamma > 0 \colon \beta'\leq\beta'' \leq \min\{1,\beta'+\gamma\} \Rightarrow \lvert F(\beta'') - F(\beta') \rvert
	\leq \frac{\delta}{2}.
\end{equation*}
It is clear that for all $k \geq K = \ceil{1/(2(\min\{1,\beta'+\gamma\}-\beta'))}$, there exist some $q\in\mathbb{N}$ such that $q/k \in [\beta', \min\{1,\beta'+\gamma\}]$ (note that since $\beta'<1$, the interval has infinite cardinality); 
as a result, $\lvert F(q/k) - F(\beta') \rvert \leq \delta/2$ for $k \geq K$.
Hence,
\begin{equation} \label{eqn:F(b')-F(q/k)<d/2}
	\forall k \geq K\colon F(\beta') - \max_{j\in\{0,\ldots,k\}} F\left(\tfrac{j}{k}\right)
	\leq F(\beta') - F\left(\tfrac{q}{k}\right)
	\stackrel{(a)}{\leq} \frac{\delta}{2},
\end{equation}
where $(a)$ follows from the fact that $x\leq|x|$ for $x\in\mathbb{R}$.
Thus, \eqref{eqn:epsilonk->0 Analytics} follows from \eqref{eqn:F(b)-F(b')<d/2} and \eqref{eqn:F(b')-F(q/k)<d/2}.

\emph{Proof of \eqref{eqn:B-Bk<r/k}}:
The first inequality follows from \eqref{eqn:B(R)} and the fact that $G^{\rm eff}_k(\cdot)$ is non-decreasing (because $G(\cdot)$ is non-increasing).
For the second inequality, define
\begin{equation} \label{eqn:barBk}
	\bar{B}_k(\rho) 
	:= \sup_{\beta\in[0,1]}{\bar{G}_k(\beta) - \rho \beta},
\end{equation}
where
$
	\bar{G}_k(\beta)
	:= G^{\rm eff}_k\left({\ceil{\beta k}}/{k} \right),
$
for $\beta\in[0,1]$.
Therefore, from \eqref{eqn:B(R)} and the fact that $G^{\rm eff}_k(\cdot)$ is non-decreasing, it can be obtained that, for all $\rho\geq 0$ and $k\in\mathbb{N}$, we have
$
	B(\rho) \leq \bar{B}_k(\rho).
$
We will prove later that there exists $i\in\{0,1,\ldots,k-1\}$ such that
\begin{equation} \label{eqn:argsupbarBk=i/k}
	\bar{B}_k(\rho)
	= G^{\rm eff}_k\left(\tfrac{i+1}{k}\right) - \rho\frac{i}{k}.
\end{equation}
As a result,
\begin{equation*}
	\bar{B}_k(\rho)
	= G^{\rm eff}_k\left(\tfrac{i+1}{k}\right) - \rho\frac{i+1}{k} + \frac{\rho}{k}
	= G_k\left(\tfrac{i+1}{k}\right) - \rho\frac{i+1}{k} + \frac{\rho}{k}
	\leq B_k(\rho) + \frac{\rho}{k}.
\end{equation*}
Thus, \eqref{eqn:B-Bk<r/k} is proved.
Now, we prove \eqref{eqn:argsupbarBk=i/k}.
There exists a sequence $\{\beta_m\}_{m=1}^\infty$ such that
\begin{equation*}
	\bar{B}_k(\rho)
	= \lim_{m\to\infty}{G_k^{\rm eff}\left(\tfrac{\ceil{\beta_m k}}{k}\right) - \rho\beta_m}.
\end{equation*}
Further, because $\beta_m\in [0,1]$ for all $m$, there exists a subsequence of $\{\beta_m\}_{m=1}^\infty$ converging to some $\hat\beta\in [0,1]$.
Without loss of generality, we assume that $\{\beta_m\}_{m=1}^\infty$, itself, converges to $\hat\beta$.
Hence, from \eqref{eqn:barBk}, we have that for all $\beta\in [0,1]$,
\begin{equation} \label{eqn:Gbm-rb>gb-r}
	\lim_{m\to\infty}{G_k^{\rm eff}\left(\tfrac{\ceil{\beta_m k}}{k}\right) - \rho\hat\beta}
	\geq G_k^{\rm eff}\left(\tfrac{\ceil{\beta k}}{k}\right) - \rho\beta.
\end{equation}
If $\rho=0$, then, we have $\beta_m=\hat\beta=1$ and $\bar{B}_k(\rho) = G_k^{\rm eff}(1)$; as a result \eqref{eqn:argsupbarBk=i/k} is true for $i=k-1$.
Otherwise, we consider three different cases
\begin{itemize}
	\item Case 1 ($\hat\beta = 1$):
	Here we have that
	\begin{equation*}
		\lim_{m\to\infty}{G_k^{\rm eff}\left(\tfrac{\ceil{\beta_m k}}{k}\right)} - \rho\hat\beta
		= G_k^{\rm eff}(1) - \rho.
	\end{equation*}
	Consider the sequence $\beta'_m = (k-1)/k + 1/m$.
	Therefore, we can write
	\begin{equation*}
		\lim_{m\to\infty}{G_k^{\rm eff}\left(\tfrac{\ceil{\beta'_m k}}{k}\right) - \rho\beta'_m}
		= G_k^{\rm eff}(1) - \rho\frac{k-1}{k}.
	\end{equation*}
	Thus, it is a contradiction with \eqref{eqn:Gbm-rb>gb-r}.
	Therefore, this case is not valid.
	
	\item Case 2 ($\hat\beta \neq i/k$ for all $i=0,1,\ldots,k$):
	Consider the sequence $\beta'_m = \floor{\hat\beta k}/k + 1/m$.
	Then, we have
	\begin{equation*}
		\lim_{m\to\infty}{G_k^{\rm eff}\left(\tfrac{\ceil{\beta'_m k}}{k}\right) - \rho\beta'_m}
		= G_k^{\rm eff}\left(\tfrac{\ceil{\hat\beta k}}{k}\right) - \rho\frac{\floor{\hat\beta k}}{k}.
	\end{equation*}
	However,
	\begin{equation*}
		\lim_{m\to\infty}{G_k^{\rm eff}\left(\tfrac{\ceil{\beta_m k}}{k}\right)} - \rho\hat\beta
		= G_k^{\rm eff}\left(\tfrac{\ceil{\hat\beta k}}{k}\right) - \rho\hat\beta.
	\end{equation*}
	This is a contradiction with \eqref{eqn:Gbm-rb>gb-r}.
	Therefore, this case is not valid.
	
	\item Case 3 ($\hat\beta = i/k$ for some $i=0,1,\ldots,k-1$):
	If $\beta_m$ tends to $i/k$, then, for large enough $m$, we have
$
		G_k^{\rm eff}\left({\ceil{\beta_m k}}/{k}\right)
		\leq G_k^{\rm eff}\left((i+1)/{k}\right).
$
	The supremum can be achieved when $\beta_m$ tends to $i/k$ from above, for example, for $\beta'_m=i/k+1/m$.
	Therefore, in this case
	\begin{equation*}
		\lim_{m\to\infty}{G_k^{\rm eff}\left(\tfrac{\ceil{\beta_m k}}{k}\right)} - \rho\hat\beta
		= G_k^{\rm eff}\left(\tfrac{i+1}{k}\right) - \rho\frac{i}{k}.
	\end{equation*}
	Thus, \eqref{eqn:argsupbarBk=i/k} is proved.
\end{itemize}
Thus, the derivation of the converse direction of the proof is complete.

\textbf{Achievability}:
Utilizing Lemma \ref{lmm:CRDFshift}, it is sufficient to prove that $\crdf$ $G^{\rm eff}$ is achievable.
We prove it in two steps:
\begin{enumerate}
	\item \label{itm:achR(D)hatG_L} $\crdf$ $\hat{G}^{\rm eff}$ satisfies the distortion constraint \eqref{eqn:limsupEd<bard},
	\item \label{itm:achR(D)G_L} $\crdf$ $G^{\rm eff}$ satisfies both the leakage constraint \eqref{eqn:mainI<L} and distortion constraint \eqref{eqn:limsupEd<bard}.
\end{enumerate}

\emph{Proof of Step \ref{itm:achR(D)hatG_L}})
For an arbitrary $k\in\mathbb{N}$, from Definition \ref{def:Code}, we have for $i=1,\ldots,k$
\begin{equation*}
	\hat{R}_i
	= \hat{G}^{\rm eff}\left( \tfrac{i}{k} \right) - \hat{G}^{\rm eff}\left( \tfrac{i-1}{k}\right)
	= \int_{\frac{i-1}{k}}^{\frac{i}{k}}{\frac{\ud \hat{G}^{\rm eff}}{\ud \alpha}(\alpha) \ud \alpha}.
\end{equation*}
From the classical rate distortion theorem \cite[Theorem 3.5]{ElGamal:2011}, we obtain that there exists a memoryless encoder for each block such that
\begin{equation} \label{inequality2}
	\limsup_{n\to\infty}\Eb{d\big(X_{(i-1)n+1}^{i n}, \hat{X}_{(i-1)n+1}^{i n} \big)}
	\leq D(k \hat{R}_i).
\end{equation}
From Remark \ref{cor:R(D)convex}, $D(\cdot)$ is convex.
So, using Jensen's inequality \cite[Theorem 2.6.2]{Cover06}, we have
\begin{equation} \label{inequality3}
	D\left(k \int_{\frac{i-1}{k}}^{\frac{i}{k}}{\frac{\ud \hat{G}^{\rm eff}}{\ud \alpha}(\alpha) \ud \alpha} \right)
	\leq k \int_{\frac{i-1}{k}}^{\frac{i}{k}}{D\left( \frac{\ud \hat{G}^{\rm eff}}{\ud \alpha}(\alpha) \right) \ud \alpha}.
\end{equation}
Therefore, from \eqref{inequality2} and \eqref{inequality3} we obtain
\begin{IEEEeqnarray*}{rCl}
	\E{d(X^n, \hat{X}^n)}
	&=& \frac{1}{k} \sum_{i=1}^k{\Eb{d\big(X_{(i-1)n+1}^{i n}, \hat{X}_{(i-1)n+1}^{i n} \big)}} \\
	&\leq& \frac{1}{k} \sum_{i=1}^k{k \int_{\frac{i-1}{k}}^{\frac{i}{k}}{D\left( \frac{\ud \hat{G}^{\rm eff}}{\ud \alpha}(\alpha) \right) \ud \alpha}}
	= \int_0^1{D\left( \frac{\ud \hat{G}^{\rm eff}}{\ud \alpha}(\alpha) \right) \ud\alpha}.
\end{IEEEeqnarray*}

\emph{Proof of Step \ref{itm:achR(D)G_L}})
From Lemma \ref{lmm:CRDFMaj}, we obtain that if $\hat{G}^{\rm eff}$ satisfies the distortion constraint \eqref{eqn:limsupEd<bard}, then there exists a sequence of coding schemes satisfying \eqref{eqn:limsupEd<bard} with $\crdf$ $G^{\rm eff}$.
Hence, utilizing Step \ref{itm:achR(D)hatG_L} of the achievability proof, we only need to show that $G^{\rm eff}$ satisfies the leakage constraint \eqref{eqn:mainI<L}.
To this end, we have that, for all $j\in\{1,\ldots,k\}$ and all $k\in\mathbb{N}$,
\begin{equation*}
	\frac{1}{nk}\Ib{X^{j k}}{M^j}
	\leq \frac{1}{nk}\Hb{M^j}
	\leq \sum_{i=1}^j R_i
	= G^{\rm eff}\left(\tfrac{j}{k}\right)
	\stackrel{(a)}{\leq} L\left(\tfrac{j}{k}\right),
\end{equation*}
where $(a)$ follows from the definition of $G^{\rm eff}$ in \eqref{eqn:GL} and the following argument:
\begin{IEEEeqnarray*}{rCl}
	G^{\rm eff}(\alpha)
	&\stackrel{(a)}{=}& \max\{0, G(\alpha) - \max_{\beta\in[0,1]}{G(\beta) - L(\beta)}\} \\
	&\leq& \max\{0, G(\alpha) - G(\alpha) + L(\alpha)\}
	= \max\{0, L(\alpha)\}
	= L(\alpha),
\end{IEEEeqnarray*}
where $(a)$ follows by selecting $\beta=\alpha$.
Thus, the achievability is derived.
\qed

\subsection{Proof of Corollary \ref{crl:Lossless=Lossy for Hamming}} \label{prf:crl:Lossless=Lossy for Hamming}
\emph{Proof of \eqref{eqn:Geff1-0>H}$\Leftrightarrow$\eqref{eqn:F1-F0<H}}:
First, we assume \eqref{eqn:Geff1-0>H} is true.
Then,
\begin{IEEEeqnarray*}{l}
	G(1) - G(\alpha)
	\stackrel{(a)}{\geq} G^{\rm eff}(1) - G^{\rm eff}(\alpha)
	\geq (1-\alpha) \H{X}, \\
	G(1) - G(\alpha)
	\stackrel{(b)}{\geq} G^{\rm eff}(1) - G^{\rm eff}(\alpha)
	\stackrel{(c)}{\geq} G^{\rm eff}(1) - L(\alpha)
	\stackrel{(d)}{\geq} \H{X} - L(\alpha),
\end{IEEEeqnarray*}
where $(a)$ and $(b)$ follow because $G^{\rm eff}(\alpha)=\max\{G(\alpha)-c,0\}$ for some $c\geq0$ and $G$ is non-decreasing;
$(c)$ follows from definition of $G^{\rm eff}$ in \eqref{eqn:GL};
$(d)$ follows by selecting $\alpha = 0$ in \eqref{eqn:Geff1-0>H} and the fact that $G^{\rm eff}(0) = 0$.
Hence, the derivation of \eqref{eqn:F1-F0<H} from \eqref{eqn:Geff1-0>H} is complete.\\
Next, we assume \eqref{eqn:F1-F0<H} is true.
First, consider the case that $G^{\rm eff}(\alpha) > 0$:
\begin{equation*}
	G^{\rm eff}(\alpha) 
	= G(\alpha)-\sup_{\beta\in[0,1]}{G(\beta) - L(\beta)}
	\Longrightarrow G^{\rm eff}(1) - G^{\rm eff}(\alpha)
	\stackrel{(a)}{=} G(1) - G(\alpha)
	\geq (1-\alpha)\H{X},
\end{equation*}
where $(a)$ follows because $G(\alpha)$ is non-decreasing.
Now, consider the case that $G^{\rm eff}(\alpha) = 0$:
\begin{IEEEeqnarray*}{l}
	G(\alpha) 
	\leq \sup_{\beta\in[0,1]}{G(\beta) - L(\beta)} \\
	\Rightarrow G^{\rm eff}(1) - G^{\rm eff}(\alpha)
	= G^{\rm eff}(1)
	\geq G(1) -\sup_{\beta\in[0,1]}{G(\beta) - L(\beta)}
	\geq G(1) -G(\alpha)
	\geq (1-\alpha) \H{X}.
\end{IEEEeqnarray*}

\emph{Proof of \eqref{eqn:intDG<d}$\Rightarrow$\eqref{eqn:Geff1-0>H}}:
When $\bar{d} = 0$, $D((\ud\hat{G}^{\rm eff}/\ud\alpha) (\alpha)) = 0$ almost everywhere due to the fact that $\hat{G}^{\rm eff}$ is differentiable everywhere except a countable number of points (see Lemma \ref{lmm:Dconvf}).
Hence, we obtain that, for all $\alpha\in(0,1)$ except a countable number of points, (straightforward extension of \cite[Theorem 10.3.1]{Cover06})
\begin{equation} \label{eqn:Geffhat'>H}
	\frac{\ud \hat{G}^{\rm eff}}{\ud \alpha}(\alpha) \geq \H{X}.
\end{equation}
Because the number of discontinuities of $\hat{G}^{\rm eff}$ is countable, \eqref{eqn:Geffhat'>H} is equivalent to
\begin{equation*}
	\hat{G}^{\rm eff}(1) - \hat{G}^{\rm eff}(\alpha)
	= \int_\alpha^1{\frac{\ud \hat{G}^{\rm eff}}{\ud \alpha}(\beta) \ud\beta} 
	\geq (1-\alpha)\H{X},
	\qquad\forall \alpha\in [0,1].
\end{equation*}
From Lemma \ref{lmm:maxf=f(b)}, we obtain that, for all $\alpha\in [0,1]$,
\begin{equation*}
	G^{\rm eff}(1) - \hat{G}^{\rm eff}(\alpha)
	\geq (1-\alpha)\H{X}
	\Longrightarrow G^{\rm eff}(1) - (1-\alpha)\H{X}
	\geq \hat{G}^{\rm eff}(\alpha).
\end{equation*}
Hence, from the definition of concave-hull, in Definition \ref{def:ConvHull}, and the fact that the function $\alpha\mapsto G^{\rm eff}(1) - (1-\alpha)\H{X}$ is linear and, as a result, concave, \eqref{eqn:Geff1-0>H} follows.
\qed

\section{Useful Lemmas} \label{sec:usefulLemmas}
In this section, we state some lemmas utilized in the proofs of our results.
\begin{lemma} \label{lmm:CRDFshift}
	Let $G_1$ and $G_2$ be two $\crdf$s such that, for $\alpha\in[0,1]$, $G_1(\alpha) = \max\{0, G_2(\alpha)-c\}$, for some $c\geq 0$.
	Then, for $k,n\in\mathbb{N}$, and a sequence of codes $(G_1, k, n)$-code$\colon x^{nk} \mapsto m_{(1)}^k \mapsto \hat{x}_{(1)}^{nk}$, there exists a sequence of codes $(G_2, k, n)$-code$\colon x^{nk} \mapsto m_{(2)}^k \mapsto \hat{x}_{(2)}^{nk}$ such that $m_{(1)}(j)=m_{(2)}(j)$ for $j=1,\ldots,k$ and for any input $x^n\in\mathcal{X}^n$ when $n$ is large enough.
\end{lemma}

\begin{IEEEproof}
	From Definition \ref{def:CRDF}, it is clear that if $G_2$ is a $\crdf$, $G_1$ is a valid $\crdf$ as well.
	According to Definition \ref{def:Code}, for $i=1,\ldots,k$, we have
$
		R_i^{(\ell)}
		= G_\ell\left( {i}/{k} \right) - G_\ell\left( (i-1)/{k}\right),
$
	for $\ell\in\{1,2\}$.
	If one shows that,
$
		R_i^{(1)} \leq R_i^{(2)},
$
	for all $i\in\{1,\ldots,k\}$, it is clear that the set of encoders and the decoder of $(G_1, k, n)$-code can be exactly used for $(G_2, k, n)$-code when $n$ is large enough;
	as a result, the lemma is proved.
	In order to prove the inequality, we consider two following cases:
	\begin{itemize}
		\item $G_1\left({i}/{k}\right) = 0$.
		Since $G_1$ is $\crdf$, it is non-decreasing; as a result, $G_1\left((i-1)/{k}\right) = 0$.
		Hence,
$
			R_i^{(1)}
			= 0
			\leq R_i^{(2)}.
$
		
		\item $G_1\left({i}/{k}\right) > 0$.
		We can write
		\begin{equation*} \textstyle
			R_i^{(1)} 
			= G_2\left(\frac{i}{k}\right) - c - G_1\left(\frac{i-1}{k}\right)
			\leq G_2\left(\frac{i}{k}\right) - c - \left(G_2\left(\frac{i-1}{k}\right) - c\right)
			= R_i^{(2)}.
		\end{equation*}
	\end{itemize}
	Therefore, the lemma is proved.
\end{IEEEproof}

\begin{lemma} \label{lmm:CRDFMaj}
	Let $G_1$ and $G_2$ be two $\crdf$s such that the following conditions hold:
	\begin{equation*}
		\begin{cases}
			G_1(\alpha) \geq G_2 (\alpha),
			\qquad \alpha\in[0,1), \\
			G_1(1) = G_2(1).
		\end{cases}
	\end{equation*}
	Then, for $k, n\in\mathbb{N}$ and a sequence of codes $(G_1, k, n)$-code$\colon x^{n k} \mapsto m_{(1)}^k \mapsto \hat{x}_{(1)}^{n k}$, there exists a sequence of codes $(G_2, k, n)$-code$\colon x^{n k} \mapsto m_{(2)}^k \mapsto \hat{x}_{(2)}^{n k}$ such that, for any $k\in\mathbb{N}$ and large enough $n$, $\hat{x}_{(1)}^{n k} = \hat{x}_{(2)}^{n k}$, for all $x^{n k}\in\mathcal{X}^{n k}$.
\end{lemma}

\begin{IEEEproof}
	For a fixed $k\in\mathbb{N}$, according to Definition \ref{def:Code}, for $i=1,\ldots,k$, $R_i^{(\ell)} = G_\ell\left( {i}/{k} \right) - G_\ell\left( (i-1)/{k}\right)$, for $\ell\in\{1,2\}$.
	Hence, we have
	\begin{equation} \label{eqn:majlmm}
		\sum_{i=1}^j{R_i^{(1)}}
		\geq\sum_{i=1}^j{R_i^{(2)}},
		\; j\in\{1,\ldots,k-1\},
		\qquad\sum_{i=1}^k{R_i^{(1)}}
		=\sum_{i=1}^k{R_i^{(2)}}.
	\end{equation}
	Later we show that there exist some $R_{i,j} \geq 0$ for $1\leq j \leq i \leq k$ such that
	\begin{equation} \label{eqn:ratesRij}
		R_j^{(1)} = \sum_{i=j}^k R_{i,j},
		\qquad R_i^{(2)} = \sum_{j=1}^i R_{i,j},
	\end{equation}
	Then, we split the message $m_{(1)}(j)$ of block $j$ with rate $R_j^{(1)}$ into messages $m(j,j),\ldots,m(k,j)$ with rates $R_{j,j},\ldots, R_{k,j}$, respectively.
	Because the coding is sequential, we can define the encoder of $(G_2, k, n)$-code for block $i$ as $m_{(2)}(i) = (m(i,1),\ldots,m(i,i))$.
	It follows because the rates of later blocks can be used in earlier blocks.
	From \eqref{eqn:ratesRij}, it is obtained that the rate of $m_{(2)}(i)$ is $R_i^{(2)}$.
	We assume that the decoder of $(G_2, k, n)$-code is the same as the decoder of $(G_1, k, n)$-code.
	Hence, it only remains to prove \eqref{eqn:ratesRij}.
	
	\emph{Proof of \eqref{eqn:ratesRij}}:
	We use induction over $k\in\mathbb{N}$.
	For $k=1$, \eqref{eqn:ratesRij} is followed because
$
		R_{1,1} = R_1^{(1)} = R_2^{(1)}. 
$
	Next, we assume that \eqref{eqn:ratesRij} is true for $k-1$, and we prove it for $k$.
	Define
	\begin{equation*}
		\begin{cases}
			\bar{R}_2^{(1)} = R_2^{(1)} + (R_1^{(1)} - R_1^{(2)}), & ~ \\
			\bar{R}_j^{(1)} = R_j^{(1)} & j\in\{3,\ldots,k\}, \\
			\bar{R}_i^{(2)} = R_i^{(2)} & i\in\{2,\ldots,k\}.
		\end{cases}
	\end{equation*}
	From \eqref{eqn:majlmm}, it is clear that 
	\begin{IEEEeqnarray*}{l}
		R_1^{(1)} - R_1^{(2)} \geq 0 \Rightarrow \bar{R}_2^{(1)} \geq 0, \\
		\sum_{i=2}^j{\bar{R}_i^{(1)}}
		= \sum_{i=1}^j{R_i^{(1)}}
		- R_1^{(2)}
		\geq \sum_{i=1}^j{R_i^{(2)}}
		- R_1^{(2)}
		=\sum_{i=2}^j{\bar{R}_i^{(2)}},
		\quad j\in\{2,\ldots,k-1\}, \\
		\sum_{i=2}^k{\bar{R}_i^{(1)}}
		= \sum_{i=1}^k{R_i^{(1)}}
		- R_1^{(2)}
		= \sum_{i=1}^k{R_i^{(2)}}
		- R_1^{(2)}
		=\sum_{i=2}^k{\bar{R}_i^{(2)}}.
	\end{IEEEeqnarray*}
	Therefore, $\{\bar{R}_j^{(1)}\}_{j=2}^k$ and $\{\bar{R}_i^{(2)}\}_{i=2}^k$ satisfy the induction assumption.
	As a result, there exists a set $\bar{R}_{i,j}$ for $2\leq j \leq i \leq k$ such that
	\begin{equation*}
		\bar{R}_j^{(1)} = \sum_{i=j}^k \bar{R}_{i,j},
		\qquad\bar{R}_i^{(2)} = \sum_{j=2}^i \bar{R}_{i,j}.
	\end{equation*}
	Now, define
	\begin{equation*}
		\begin{cases}
			R_{1,1} = R_1^{(2)}, & ~ \\
			R_{i,1} = \frac{R_1^{(1)}-R_1^{(2)}}{R_2^{(1)} + R_1^{(1)}-R_1^{(2)}} \bar{R}_{i,2},
			\quad R_{i,2} = \frac{R_2^{(1)}}{R_2^{(1)} + R_1^{(1)}-R_1^{(2)}} \bar{R}_{i,2}, & i\in\{2,\ldots,k\}, \\
			R_{i,j} = \bar{R}_{i,j}, & 3 \leq j \leq i \leq k.
		\end{cases}
	\end{equation*}
	To prove \eqref{eqn:ratesRij}, we can write
	\begin{IEEEeqnarray*}{rCl}
		\sum_{i=1}^k R_{i,1}
		&=& R_1^{(2)} + \tfrac{R_1^{(1)}-R_1^{(2)}}{R_2^{(1)} + R_1^{(1)}-R_1^{(2)}} \sum_{i=2}^k \bar{R}_{i,2}
		= R_1^{(2)} + \tfrac{R_1^{(1)}-R_1^{(2)}}{R_2^{(1)} + R_1^{(1)}-R_1^{(2)}} \bar{R}_2^{(1)} \\
		&=& R_1^{(2)} + \tfrac{R_1^{(1)}-R_1^{(2)}}{R_2^{(1)} + R_1^{(1)}-R_1^{(2)}} (R_2^{(1)} + (R_1^{(1)} - R_1^{(2)}))
		= R_1^{(1)}, \\
		\sum_{i=2}^k R_{i,2}
		&=& \tfrac{R_2^{(1)}}{R_2^{(1)} + R_1^{(1)}-R_1^{(2)}} \sum_{i=2}^k \bar{R}_{i,2}
		= \tfrac{R_2^{(1)}}{R_2^{(1)} + R_1^{(1)}-R_1^{(2)}} \bar{R}_2^{(1)} 
		= \tfrac{R_2^{(1)}}{R_2^{(1)} + R_1^{(1)}-R_1^{(2)}} (R_2^{(1)} + (R_1^{(1)} - R_1^{(2)}))
		= R_2^{(1)}, \\
		\sum_{i=j}^k R_{i,j}
		&=& \sum_{i=j}^k \bar{R}_{i,2}
		= \bar{R}_j^{(1)}
		= R_j^{(1)}, 
		\qquad j\in\{3,\ldots,k\}, \\
		\sum_{j=1}^1 R_{i,j}
		&=& R_{1,1}
		= R_1^{(2)}, \\
		\sum_{j=1}^2 R_{i,j}
		&=& R_{2,1} + R_{2,2}
		= \tfrac{R_1^{(1)}-R_1^{(2)}}{R_2^{(1)} + R_1^{(1)}-R_1^{(2)}} \bar{R}_{2,2}
		+ \tfrac{R_2^{(1)}}{R_2^{(1)} + R_1^{(1)}-R_1^{(2)}} \bar{R}_{2,2}
		= \bar{R}_{2,2}
		= \bar{R}_2^{(2)}
		= R_2^{(2)}, \\
		\sum_{j=1}^i R_{i,j}
		&=& R_{2,1} + R_{2,2} 
		+ \sum_{j=3}^i \bar{R}_{i,j}
		= \bar{R}_{i,2}
		+ \sum_{j=3}^i \bar{R}_{i,j}
		= \sum_{j=2}^i \bar{R}_{i,j}
		= \bar{R}_2^{(2)}
		= R_2^{(2)},
		\quad i\in\{3,\ldots,k\}.
	\end{IEEEeqnarray*}
	Thus, \eqref{eqn:ratesRij} and, as a result, the lemma is proved.
\end{IEEEproof}

\begin{lemma} \label{lmm:maxf=f(b)}
	Let $f \colon [a,b] \to \mathbb{R}$ be non-decreasing where $\hat{f}$ denotes the envelope of $f$.
	Then,\\
	\begin{inparaenum}
		\item \label{itm:bf(b)=f(b)}
		$\hat{f}(b) = f(b)$,\qquad
		\item \label{itm:bf(a)=f(a)}
		$\hat{f}(a) = f(a)$.
	\end{inparaenum}
\end{lemma}

\begin{IEEEproof}
	\emph{Proof of \ref{itm:bf(b)=f(b)})}:
	Due to the definition of the concave-hull, $\hat{f}(b) \geq f(b)$.
	It cannot be strictly greater than $f(b)$ because the function $x \mapsto f(b) $, for $x\in [a, b]$, is concave and always greater than or equal to $f(x)$ because $f$ is non-decreasing.
	However, the function is not always greater than $\hat{f}(x)$, which is a contradiction.
	Hence, $\hat{f}(b) = f(b)$.
	
	\emph{Proof of \ref{itm:bf(a)=f(a)})}:
	From the definition of the envelope, we have $\hat{f}(a) \geq f(a)$.
	We assume $\hat{f}(a) > f(a)$ and define the function $\hat{f}'(x)$ as $\hat{f}(x)$ for $x\in(a,b]$ and $f(a)$ for $x=a$.
	As a result, $f(x) \leq \hat{f}'(x) \leq \hat{f}(x)$ for $x\in[a,b]$.
	Hence, if we show that $\hat{f}'(x)$ is concave, it will be a contradiction, and the result follows.
	We need to show that for all $a \leq x_1 < x_2 \leq b$ and $t\in (0,1)$ we have
	\begin{equation} \label{eqn:hatf' concave}
		\hat{f}'(t x_1 + (1-t) x_2) \geq t \hat{f}'(x_1) + (1-t) \hat{f}'(x_2).
	\end{equation}
	If $x_1>a$, we have $\hat{f}'(t x_1 + (1-t) x_2) = \hat{f}(t x_1 + (1-t) x_2)$, $\hat{f}'(x_1) = \hat{f}(x_1)$, and $\hat{f}'(x_2) = \hat{f}(x_2)$.
	Hence, \eqref{eqn:hatf' concave} follows from the concavity of $\hat{f}$.
	If $x_1=a$, we have $\hat{f}'(t x_1 + (1-t) x_2) = \hat{f}(t x_1 + (1-t) x_2)$, $\hat{f}'(x_1) < \hat{f}(x_1)$, and $\hat{f}'(x_2) = \hat{f}(x_2)$.
	Therefore, we obtain
	\begin{equation*}
		\hat{f}'(t x_1 + (1-t) x_2)
		= \hat{f}(t x_1 + (1-t) x_2) 
		\stackrel{(a)}{\geq} t \hat{f}(x_1) + (1-t) \hat{f}(x_2)
		\geq t \hat{f}'(x_1) + (1-t) \hat{f}'(x_2),
	\end{equation*}
	where $(a)$ follows from the concavity of $\hat{f}$.
	Thus, $\hat{f}'$ is concave.
\end{IEEEproof}

\begin{lemma}[Majorization Inequality] \cite[p. 14]{Marshall10} \label{lmm:MajIneq}
	Consider two sequences $x_1\geq\cdots\geq x_k$ and $y_1 \geq \cdots \geq y_k$ such that $\{x_i\}_{i=1}^k$ majorizes $\{y_i\}_{i=1}^k$, i.e.,
	\begin{IEEEeqnarray*}{l}
		\begin{cases}
		\sum_{i=1}^j {x_i} \geq \sum_{i=1}^j {y_i},
		& \forall j=1,\ldots,k-1, \\
		\sum_{i=1}^k {x_i} = \sum_{i=1}^k {y_i}.
		\end{cases}
	\end{IEEEeqnarray*}
	Then, for any convex function $f\colon \mathbb{R}\to\mathbb{R}$ we have
	\begin{equation*}
		\sum_{i=1}^k{f(x_i)}
		\geq \sum_{i=1}^k{f(y_i)}.
	\end{equation*}
\end{lemma}

\begin{lemma} \label{lmm:Dconvf}
	Let $f \colon [0,1] \to [0,\infty)$ be non-decreasing and bounded.
	The envelope of $f$ (see Definition \ref{def:ConvHull}), $\hat{f}$, for $x\in (0,1]$, is 
	\begin{equation} \label{eqn:fconv=minimax}
		\hat{f}(x) 
		= \min_{a\geq 0}\sup_{z\in[0,1]}{f(z) - a(z - x)}.
	\end{equation}
	Further, $\hat{f}$ is differentiable over $(0,1)$, except for a countable number of points, and the derivative, for $x\in (0,1)$, is
	\begin{equation} \label{eqn:Dfconv=argminimax}
		\frac{\ud f}{\ud x}(x) 
		= \mathrm{arg}\min_{a \geq 0} \sup_{z\in[0,1]}{f(z) - a (z - x)}.
	\end{equation}
	Moreover, the derivarive at point $x$ is in the interval $[0,f(1)/x]$.
\end{lemma}

\begin{IEEEproof}
	We believe that the proof exists somewhere in the literature, but we did not find it.
	So, to make the paper self-contained, we prove it again.
	Before starting to prove the lemma, we prove that \eqref{eqn:fconv=minimax} and \eqref{eqn:Dfconv=argminimax} are well-defined, i.e., the minimum is achievable:
	\begin{equation} \label{eqn:inf=min}
		\inf_{a\geq 0}\sup_{z\in[0,1]}{f(z) - a(z - x)}
		= \min_{a\geq 0}\sup_{z\in[0,1]}{f(z) - a(z - x)}.
	\end{equation}
	
	\textbf{Proof of \eqref{eqn:inf=min}}:
	Define
	\begin{equation} \label{eqn:b(a)}
		b(a) := \sup_{z\in[0,1]}{f(z) - a z}.
	\end{equation}
	Then, from \eqref{eqn:fconv=minimax}, we must show that
	\begin{equation} \label{eqn:f=min ax+b}
		\hat{f}(x)
		= \min_{a\geq 0}{a x + b(a)}
		= \inf_{a\geq 0}{a x + b(a)}.
	\end{equation}
	Later, we prove that
	\begin{equation} \label{eqn:f0<ba<f1}
		b(a) \text{ is convex},
		\qquad 0 \leq b(a) \leq f(1).
	\end{equation}
	Therefore, for any $x\in(0,1]$, if $a > f(1) / x$, we obtain that
$
		a x + b(a) > f(1) \geq b(0) = a. 0 + b(0).
$
	Hence, in order to find the infimum of $a x + b(a)$ over $a\in [0,\infty)$, we only need to consider $a\in [0 , f(1) / x]$, i.e., for $x\in(0,1]$, we have
$
		\inf_{a\geq 0}{a x + b(a)}
		= \inf_{a\in [0,f(1) / x]}{a x + b(a)}.
$

	Utilizing \eqref{eqn:f0<ba<f1}, the function $(0,f(1) / x) \to [0,\infty), \quad a \mapsto a x + b(a)$ is convex and bounded; thus, continuous and bounded.
	Hence, it has a minimum in the interval $[0,f(1) / x]$, i.e., the infimum is achievable.
	As a result, the expressions \eqref{eqn:fconv=minimax} and \eqref{eqn:Dfconv=argminimax} are well-defined.
	Hence, it remains to prove \eqref{eqn:f0<ba<f1}.

	\emph{Proof of \eqref{eqn:f0<ba<f1}}:
	For $t \in [0,1]$ and $a_1, a_2 \geq 0$, we can write
	\begin{IEEEeqnarray*}{rCl}
		b(t a_1 + (1-t) a_2)
		\!&=&\!\! \sup_{z\in[0,1]}{f(z) - (t a_1 + (1-t) a_2) z}
		\!=\!\! \sup_{z\in[0,1]}{t(f(z) - a_1 z) + (1-t)(f(z) - a_2 z)} \\
		&\stackrel{(a)}{\leq}& \sup_{z\in[0,1]}{t(f(z) - a_1 z)}
		+ \sup_{z\in[0,1]}{(1-t)(f(z) - a_2 z)}
		= t b(a_1) + (1-t) b(a_2),
	\end{IEEEeqnarray*}
	where $(a)$ follows from the fact that for any functions $f(x)$ and $g(x)$, we have $\sup_x{f(x) + g(x)} \leq \sup_x{f(x)} + \sup_x{g(x)}$.
	Hence, $b(a)$ is convex.
	Next, note that, for all $a \geq 0$,
	\begin{equation*}
		0 \leq f(0) = f(0) - a . 0 \leq b(a) \leq \sup_{z\in[0,1]} f(z) 
		\stackrel{(a)}{=} f(1),
	\end{equation*}
	where $(a)$ follows from because $f$ is non-deccreasing.
	Thus, the proof of \eqref{eqn:f0<ba<f1} is complete.
	
	\textbf{Proof of \eqref{eqn:fconv=minimax}}:
	Recalling Definition \ref{def:ConvHull}, we must prove
	
	\begin{inparaenum}
		\item \label{eqn:convf is conv}
		$\hat{f}(x) \text{ is concave}$,
		\quad\item \label{eqn:convf>f}
		$\hat{f}(x) \geq f(x), \; \forall x\in[0,1]$,
		\quad\item \label{eqn:convf>g}
		$\hat{f}(x) \leq g(x)$, for all concave $g(x)$ such that $g(x) \geq f(x), \; \forall x\in[0,1]$.
	\end{inparaenum}
	
	\emph{Proof of \ref{eqn:convf is conv}})
	Let $t \in [0,1]$ and $x_1, x_2 \in \mathcal{A}$.
	Therefore, from \eqref{eqn:f=min ax+b}, we obtain that
	\begin{IEEEeqnarray*}{rCl}
		\hat{f}(t x_1 + (1-t) x_2)
		&=& \min_{a\geq 0}{a(t x_1 + (1-t) x_2) + b(a)}
		= \min_{a\geq 0}{t (a x_1+ b(a)) + (1-t)(a x_2 + b(a))} \\
		&\geq& \min_{a\geq 0}{t (a x_1+ b(a))}
		+ \min_{a\geq 0}{(1-t)(a x_2 + b(a))} \label{eqn:fconvisconv1}
		= t \hat{f}(x_1) + (1-t) \hat{f}(x_2),
	\end{IEEEeqnarray*}
	where the inequality follows from the fact that for any functions $f(x)$ and $g(x)$, we have $\min_x{f(x) + g(x)} \geq \min_x{f(x)} + \min_x{g(x)}$.
	
	\emph{Proof of \ref{eqn:convf>f}})
	From \eqref{eqn:f=min ax+b}, we have
	\begin{equation*}
		\hat{f}(x) = \min_{a\geq 0}{ax + b(a)}
		\stackrel{(a)}{\geq} \min_{a\geq 0}{ax + f(x) - ax}
		= f(x),
	\end{equation*}
	where $(a)$ follows from \eqref{eqn:b(a)}, by substituting $z=x$.

	\emph{Proof of \ref{eqn:convf>g}})
	Assume that there exists a concave function $g(x)$ such that $g(x)\geq f(x)$ for all $x\in[0,1]$ and there exists $x_0\in [0,1]$ such that $g(x_0) < \hat{f}(x_0)$.
	From the supporting hyper plane theorem \cite[p. 51]{Boyd04}, we obtain that there exists $\hat{a}\in\mathbb{R}$ such that $g(x) \leq g(x_0) + \hat{a}(x - x_0)$ for all $x \in [0,1]$.
	As a result, $f(x) \leq g(x_0) + \hat{a}(x - x_0)$ for all $x \in [0,1]$.
	Note that since $f(x)$ is an increasing function, we have that $f(x) \leq g(x_0)$ for all $x \in [0,1]$ if $\hat{a} < 0$, which is similar to the case that we consider $\hat{a} = 0$.
	Therefore, we can assume that there exists $\hat{a} \geq 0$ such that $f(x) \leq g(x_0) + \hat{a}(x - x_0)$ for all $x \in [0,1]$.
	Hence,
	\begin{equation*}
		f(x) \leq g(x_0) + \hat{a}(x - x_0)
		\Rightarrow f(x) - \hat{a} x \leq g(x_0) - \hat{a} x_0
		\stackrel{(a)}{\Rightarrow} b(\hat{a}) \leq g(x_0) - \hat{a} x_0
		\stackrel{(b)}{\Rightarrow} \hat{a} x_0 + b(\hat{a}) < \hat{f}(x_0),
	\end{equation*}
	where $(a)$ is true because of \eqref{eqn:b(a)};
	$(b)$ is true because of the assumption $g(x_0) < \hat{f}(x_0)$.
	Thus, it contradicts with \eqref{eqn:f=min ax+b}; as a result $g(x) \geq \hat{f}(x)$ for all $x\in [0,1]$.
	
	\textbf{Proof of \eqref{eqn:Dfconv=argminimax}}:
	Define
	\begin{equation} \label{eqn:a*x0}
		a^*(x_0) 
		:= \mathrm{arg}\min_{a \geq 0} \sup_{z\in[0,1]}{f(z) - a (z - x_0)}
		= \mathrm{arg}\min_{a \geq 0} {a x_0 + b(a)}.
	\end{equation}
	If we show that, for all $x\in [0,1]$,
	\begin{equation} \label{eqn:a*=subg}
		\hat{f}(x) \leq \hat{f}(x_0) + a^*(x_0)(x - x_0),
	\end{equation}
	then, we have proved that $a^*(x_0)$ is the subgradient of $\hat{f}(x)$ at $x=x_0$ \cite[p. 338]{Boyd04}; as a result $a^*(x_0)$ is the derivative of $\hat{f}(x)$ at point $x=x_0$ if $\hat{f}(x)$ is differentiable at that point.
	It is known that convex and bounded functions defined over a compact set, have a countable number of non-differentiable points.
	Hence, \eqref{eqn:Dfconv=argminimax} is valid for all $x\in[0,1]$ except a countable number of points.
	Therefore, it only remains to prove \eqref{eqn:a*=subg}.
	From \eqref{eqn:fconv=minimax}, we know that
$
		\hat{f}(x) 
		= a^*(x) x + b(a^*(x)) \leq a^*(x_0) x + b(a^*(x_0)).
$
	Hence,
$
		\hat{f}(x)
		\leq a^*(x_0) x_0 + b(a^*(x_0)) + a^*(x_0) (x-x_0)
		= \hat{f}(x_0) + a^*(x_0) (x-x_0),
$
	where the last equality is due to \eqref{eqn:a*x0}.
	This completes the proof.
\end{IEEEproof}

\begin{lemma} \label{lmm:Linear R(D)}
	Assume a sequence of $\IID$ $\rv$s with rate-distortion function 
	\begin{equation} \label{eqn:R(D) linear 1}
		R(D) = 
		\begin{cases}
			c - D & 0\leq D \leq c, \\
			0 & D>c,
		\end{cases}
	\end{equation}
	for some $c>0$.
	Then, a $\crdf$ $G$ is achievable, given $\cldf$ $L$, with distortion $\bar{d}$, if and only if
	\begin{equation*}
		G^\mathrm{eff}(1) - G^\mathrm{eff}(\alpha)
		\geq (1 - \alpha) c - \bar{d},
		\qquad\forall \alpha\in \left[0, 1 - \frac{\bar{d}}{c}\right]
	\end{equation*}
\end{lemma}
\begin{IEEEproof}
	From \eqref{eqn:R(D) linear 1} we obtain that
	\begin{equation} \label{eqn:D(R) linear 1}
		D(R) = 
		\begin{cases}
			c - R & 0\leq R \leq c, \\
			0 & R>c.
		\end{cases}
	\end{equation}
	Later, we show that, for 
	\begin{equation} \label{eqn:Geff(beta) subgrad}
	\beta = \argmax_{0\leq\alpha\leq 1}{\hat{G}^\mathrm{eff}(\alpha) - \alpha c},
	\end{equation}
	we have
	\begin{equation} \label{eqn:Geff'<1>1}
		\begin{cases}
			\frac{\ud}{\ud\alpha}\hat{G}^\mathrm{eff}(\alpha) \geq c & \alpha<\beta, \\
			\frac{\ud}{\ud\alpha}\hat{G}^\mathrm{eff}(\alpha) \leq c & \alpha>\beta.
		\end{cases}
	\end{equation}
	Hence, we have that
	\begin{equation*}
		\int_0^1 {D\left(\frac{\ud \hat{G}^{\rm eff}}{\ud \alpha}(\alpha) \right) \ud \alpha}
		\stackrel{(a)}{=} \int_\beta^1 {c - \frac{\ud \hat{G}^{\rm eff}}{\ud \alpha}(\alpha) \ud \alpha}
		= c(1 - \beta) - \hat{G}^{\rm eff}(1) + \hat{G}^{\rm eff}(\beta),
	\end{equation*}
	where $(a)$ follows from \eqref{eqn:Geff'<1>1} and $D(R)$ in \eqref{eqn:D(R) linear 1}.
	From \eqref{eqn:Geff(beta) subgrad}, we obtain that
	\begin{equation*}
		\int_0^1 {D\left(\frac{\ud \hat{G}^{\rm eff}}{\ud \alpha}(\alpha) \right) \ud \alpha}
		= c - \hat{G}^\mathrm{eff}(1) + \max_{\alpha\in [0,1]}{\hat{G}^\mathrm{eff}(\alpha) - c\alpha}.
	\end{equation*}
	Thus, from Theorem \ref{thm:R(D)}, $G$ is achievable, given $L$, with distortion $\bar{d}$, if and only if
	\begin{equation*}
		c - \hat{G}^\mathrm{eff}(1) + \hat{G}^\mathrm{eff}(\alpha) - c \alpha
		\leq\bar{d},
		\qquad \forall\alpha\in [0,1].
	\end{equation*}
	Because the function $\alpha \mapsto c \alpha + \hat{G}^\mathrm{eff}(1) + \bar{d} - c$ is concave, from the definition of concave-hull (see Definition \ref{def:ConvHull}) we obtain that it is equivalant to
	\begin{equation*}
		G^\mathrm{eff}(\alpha)
		\leq c\alpha + \hat{G}^\mathrm{eff}(1) + \bar{d} - c 
		\stackrel{(a)}{=} G^\mathrm{eff}(1) + \bar{d} - (1-\alpha)c,
		\qquad \forall \alpha\in [0,1],
	\end{equation*}
	where $(a)$ follows from Lemma \ref{lmm:maxf=f(b)}.
	Note that it is already valid for $\alpha\in[1-\bar{d}/c, 1]$ because $G^\mathrm{eff}$ is non-decreasing; as a result $G^\mathrm{eff}(\alpha)\leq G^\mathrm{eff}(1)$.
	Therefore, it only remains to prove \eqref{eqn:Geff'<1>1}.

	\emph{Proof of \eqref{eqn:Geff'<1>1}}:
	Since $\hat{G}^\mathrm{eff}$ is continuous and bounded, it has a maximum and $\beta$ is well-defined.
	We define
$
		b(\gamma) 
		:= \sup_{0\leq\alpha\leq 1}{\hat{G}^\mathrm{eff}(\alpha) - \gamma\alpha},
$
	for $0<\gamma<1$.
	From \eqref{eqn:Dfconv=argminimax} in Lemma \ref{lmm:Dconvf}, we have,
	\begin{equation*}
		\frac{\ud}{\ud\alpha}\hat{G}^\mathrm{eff}(\alpha) \alpha 
		+ b\left(\frac{\ud}{\ud\alpha}\hat{G}^\mathrm{eff}(\alpha)\right)
		\leq c \alpha + b(c).
	\end{equation*}
	Further, from \eqref{eqn:Geff(beta) subgrad}, we can write
	\begin{IEEEeqnarray*}{l}
		c \beta + b(c)
		= \hat{G}^\mathrm{eff}(\beta)
		\leq \hat{G}^\mathrm{eff}(\beta) - \frac{\ud}{\ud\alpha}\hat{G}^\mathrm{eff}(\alpha) \beta
		+ \frac{\ud}{\ud\alpha}\hat{G}^\mathrm{eff}(\alpha) \beta
		\leq b\left(\frac{\ud}{\ud\alpha}\hat{G}^\mathrm{eff}(\alpha)\right)
		+ \frac{\ud}{\ud\alpha}\hat{G}^\mathrm{eff}(\alpha) \beta.
	\end{IEEEeqnarray*}
	Now, by adding the inequalities, we obtain
	\begin{equation*}
		\frac{\ud}{\ud\alpha}\hat{G}^\mathrm{eff}(\alpha) \alpha
		+ c\beta
		\leq c\alpha 
		+ \frac{\ud}{\ud\alpha}\hat{G}^\mathrm{eff}(\alpha) \beta
		\Longrightarrow
		\left(\frac{\ud}{\ud\alpha}\hat{G}^\mathrm{eff}(\alpha) - c \right)
		\left(\alpha-\beta\right)
		\leq 0.
	\end{equation*}
	This completes the proof.
\end{IEEEproof}

\section{Conclusions} \label{sec:conclusion}
In this work, we introduced the concept of achievable $\crdf$ to characterize the rate profiles of the sequential encoding processes that ensure a secure lossless or lossy reconstruction subject to a fidelity criterion using a joint decoder.
For $\IID$ sources, we derived a necessary and sufficient condition on the $\crdf$ for a given $\IID$ source, which is characterized by the concave-hull of the $\crdf$.
Further, we studied the case including a security constraint.
The information leakage was defined sequentially based on the mutual information between the source and its compressed representation, as it evolves. 
To characterize the security constraints, we introduced the concept of $\cldf$, which determines the allowed information leakage as distributed over encoded sub-blocks.
Finally, we derived a necessary and sufficient condition on the achievable $\crdf$ for a given $\IID$ source and $\cldf$.
We showed that the concave-hull of the effective $\crdf$, which is the amount of $\crdf$ used in the compression, characterizes the optimal achievable rate distribution.

\bibliographystyle{IEEEtran}
\bibliography{IEEEabrv,myref}

\begin{thebibliography}{10}
\providecommand{\url}[1]{#1}
\csname url@samestyle\endcsname
\providecommand{\newblock}{\relax}
\providecommand{\bibinfo}[2]{#2}
\providecommand{\BIBentrySTDinterwordspacing}{\spaceskip=0pt\relax}
\providecommand{\BIBentryALTinterwordstretchfactor}{4}
\providecommand{\BIBentryALTinterwordspacing}{\spaceskip=\fontdimen2\font plus
\BIBentryALTinterwordstretchfactor\fontdimen3\font minus
  \fontdimen4\font\relax}
\providecommand{\BIBforeignlanguage}[2]{{%
\expandafter\ifx\csname l@#1\endcsname\relax
\typeout{** WARNING: IEEEtran.bst: No hyphenation pattern has been}%
\typeout{** loaded for the language `#1'. Using the pattern for}%
\typeout{** the default language instead.}%
\else
\language=\csname l@#1\endcsname
\fi
#2}}
\providecommand{\BIBdecl}{\relax}
\BIBdecl

\bibitem{hamid19}
H.~{Ghourchian}, P.~A. {Stavrou}, T.~J. {Oechtering}, and M.~{Skoglund},
  ``Block source coding with sequential encoding,'' in \emph{IEEE Information
  Theory Workshop (ITW)}, 2019, pp. 1--5.

\bibitem{Shannon59}
C.~E. Shannon, ``Coding theorems for a discrete source with a fidelity
  criterion,'' \emph{IRE Nat. Conv. Rec}, vol.~4, no.~1, pp. 325--350, 1959.

\bibitem{berger:1971}
T.~Berger, \emph{Rate Distortion Theory: A Mathematical Basis for Data
  Compression}.\hskip 1em plus 0.5em minus 0.4em\relax Englewood Cliffs, NJ:
  Prentice-Hall, 1971.

\bibitem{Neuhoff82}
D.~Neuhoff and R.~Gilbert, ``Causal source codes,'' \emph{{IEEE} Trans. Inf.
  Theory}, vol.~28, no.~5, pp. 701--713, 1982.

\bibitem{Weissman05}
T.~{Weissman} and N.~{Merhav}, ``On causal source codes with side
  information,'' \emph{{IEEE} Trans. Inf. Theory}, vol.~51, no.~11, pp.
  4003--4013, 2005.

\bibitem{Linder01}
T.~Linder and G.~Lagosi, ``A zero-delay sequential scheme for lossy coding of
  individual sequences,'' \emph{{IEEE} Trans. Inf. Theory}, vol.~47, no.~6, pp.
  2533--2538, 2001.

\bibitem{Stavrou18}
P.~A. Stavrou, J.~{\O}stergaard, and C.~D. Charalambous, ``Zero-delay rate
  distortion via filtering for vector-valued {Gaussian} sources,'' \emph{IEEE
  J. Sel. Topics Signal Process.}, vol.~12, no.~5, pp. 841--856, 2018.

\bibitem{Tanaka18}
T.~{Tanaka}, P.~M. {Esfahani}, and S.~K. {Mitter}, ``{LQG} control with minimum
  directed information: Semidefinite programming approach,'' \emph{{IEEE}
  Trans. Autom. Control}, vol.~63, no.~1, pp. 37--52, 2018.

\bibitem{Akyol14}
E.~Akyol, K.~B. Viswanatha, K.~Rose, and T.~A. Ramstad, ``On zero-delay
  source-channel coding,'' \emph{{IEEE} Trans. Inf. Theory}, vol.~60, no.~12,
  pp. 7473--7489, 2014.

\bibitem{Matloub06}
S.~Matloub and T.~Weissman, ``Universal zero-delay joint source--channel
  coding,'' \emph{{IEEE} Trans. Inf. Theory}, vol.~52, no.~12, pp. 5240--5250,
  2006.

\bibitem{Merhav03}
N.~Merhav and I.~Kontoyiannis, ``Source coding exponents for zero-delay coding
  with finite memory,'' \emph{{IEEE} Trans. Inf. Theory}, vol.~49, no.~3, pp.
  609--625, 2003.

\bibitem{Viswanathan00}
H.~Viswanathan and T.~Berger, ``Sequential coding of correlated sources,''
  \emph{{IEEE} Trans. Inf. Theory}, vol.~46, no.~1, pp. 236--246, 2000.

\bibitem{Ma11}
N.~Ma and P.~Ishwar, ``On delayed sequential coding of correlated sources,''
  \emph{{IEEE} Trans. Inf. Theory}, vol.~57, no.~6, pp. 3763--3782, 2011.

\bibitem{Shannon49}
C.~E. Shannon, ``Communication theory of secrecy systems,'' \emph{Bell Sys.
  Tech. J.}, vol.~28, no.~4, pp. 656--715, 1949.

\bibitem{Wyner75}
A.~D. {Wyner}, ``The wire-tap channel,'' \emph{Bell Sys. Tech. J.}, vol.~54,
  no.~8, pp. 1355--1387, 1975.

\bibitem{Yamamoto97}
H.~{Yamamoto}, ``Rate-distortion theory for the {Shannon} cipher system,''
  \emph{{IEEE} Trans. Inf. Theory}, vol.~43, no.~3, pp. 827--835, 1997.

\bibitem{Prabhakaran07}
V.~{Prabhakaran} and K.~{Ramchandran}, ``On secure distributed source coding,''
  in \emph{2007 IEEE Information Theory Workshop}, 2007, pp. 442--447.

\bibitem{Gunduz08}
D.~{Gunduz}, E.~{Erkip}, and H.~V. {Poor}, ``Secure lossless compression with
  side information,'' in \emph{2008 IEEE Information Theory Workshop}, 2008,
  pp. 169--173.

\bibitem{Villard13}
J.~{Villard} and P.~{Piantanida}, ``Secure multiterminal source coding with
  side information at the eavesdropper,'' \emph{{IEEE} Trans. Inf. Theory},
  vol.~59, no.~6, pp. 3668--3692, 2013.

\bibitem{Schieler14}
C.~{Schieler} and P.~{Cuff}, ``Rate-distortion theory for secrecy systems,''
  \emph{{IEEE} Trans. Inf. Theory}, vol.~60, no.~12, pp. 7584--7605, 2014.

\bibitem{Kaspi15}
Y.~{Kaspi} and N.~{Merhav}, ``Zero-delay and causal secure source coding,''
  \emph{{IEEE} Trans. Inf. Theory}, vol.~61, no.~11, pp. 6238--6250, 2015.

\bibitem{Shamai03}
S.~Shamai and A.~Steiner, ``A broadcast approach for a single-user slowly
  fading {MIMO} channel,'' \emph{{IEEE} Trans. Inf. Theory}, vol.~49, no.~10,
  pp. 2617--2635, 2003.

\bibitem{Jorswieck07}
E.~Jorswieck and H.~Boche, ``Majorization and matrix-monotone functions in
  wireless communications,'' \emph{Foundations and Trends in Communications and
  Information Theory}, vol.~3, no.~6, pp. 553--701, 2007.

\bibitem{Palomar07}
D.~P. Palomar and Y.~Jiang, ``{MIMO} transceiver design via majorization
  theory,'' \emph{Foundations and Trends in Communications and Information
  Theory}, vol.~3, no. 4-5, pp. 331--551, 2007.

\bibitem{ElGamal:2011}
A.~El~Gamal and Y.-H. Kim, \emph{Network Information Theory}.\hskip 1em plus
  0.5em minus 0.4em\relax Cambridge University press, 2011.

\bibitem{Boyd04}
S.~Boyd and L.~Vandenberghe, \emph{Convex Optimization}.\hskip 1em plus 0.5em
  minus 0.4em\relax Cambridge university press, 2004.

\bibitem{Cover06}
T.~M. Cover and J.~A. Thomas, \emph{Elements of Information Theory},
  2nd~ed.\hskip 1em plus 0.5em minus 0.4em\relax New York: John Wiley {\&}
  Sons, 2006.

\bibitem{Courtade13}
T.~A. Courtade and T.~Weissman, ``Multiterminal source coding under logarithmic
  loss,'' \emph{{IEEE} Trans. Inf. Theory}, vol.~60, no.~1, pp. 740--761, 2013.

\bibitem{Shkel17}
Y.~Shkel, M.~Raginsky, and S.~Verd{\'u}, ``Universal lossy compression under
  logarithmic loss,'' in \emph{International Symposium on Information Theory
  (ISIT)}.\hskip 1em plus 0.5em minus 0.4em\relax IEEE, 2017, pp. 1157--1161.

\bibitem{Gray09}
R.~M. Gray, \emph{Probability, Random Processes, and Ergodic Properties}.\hskip
  1em plus 0.5em minus 0.4em\relax Springer, 2009.

\bibitem{Marshall10}
A.~W. Marshall, I.~Olkin, and B.~C. Arnold, \emph{Inequalities: Theory of
  Majorization and Its Applications}, 2nd~ed.\hskip 1em plus 0.5em minus
  0.4em\relax Springer Science \& Business Media, 2010.

\end{thebibliography}

\end{document}